\documentclass[twocol]{IEEEtran}
\usepackage{amsmath,amsfonts}
\usepackage{algpseudocode}
\usepackage{algorithm}
\usepackage{array}
\usepackage[caption=false,font=normalsize,labelfont=sf,textfont=sf]{subfig}
\usepackage{textcomp}
\usepackage{stfloats}
\usepackage{url}
\usepackage{verbatim}
\usepackage{graphicx}
\usepackage{cite}
\IEEEoverridecommandlockouts
\usepackage{amssymb}
\usepackage{balance}
\usepackage[justification=centering]{caption}
\usepackage{booktabs}
\usepackage{tabto}
\usepackage{multirow}
\usepackage[dvipsnames]{xcolor}

\begin{document}

\title{
Vol-Mark: A Watermark for 3D Medical Volume Data Via Cubic Difference Expansion and Contrastive Learning
\\[0.2cm]
}

\author{Jiangnan Zhu, Yuntao Wang, Shengli Pan,  Yujie Gu
\thanks{J. Zhu and Y. Gu are with Kyushu University, Fukuoka, Japan. 
Y. Wang is with The University of Electro-Communications, Tokyo, Japan.
S. Pan is with Beijing University of Posts and Telecommunications, Beijing, China.
(e-mails: zhu.jiangnan.584@s.kyushu-u.ac.jp; y-wang@uec.ac.jp; psl@bupt.edu.cn; gu@inf.kyushu-u.ac.jp.)
J. Zhu was supported in part by the WISE program (MEXT) at Kyushu University in this work.
}
}

\maketitle

\begin{abstract}
Today, advances in medical technology extensively utilize 3D volume data for accurate and efficient diagnostics.
However, sharing these data across networks in telemedicine poses significant security risks of data tampering and unauthorized copying.
To address these challenges, this paper proposes a novel reversible-zero watermarking approach, termed Vol-Mark, for medical volume data to protect their ownership and authenticity in telemedicine.
The proposed Vol-Mark method offers two key benefits:
1) it designs a volume data feature extractor that leverages contrastive learning to efficiently extract discriminative and stable volumetric features, ensuring robustness against 3D attacks;
2) it introduces the cubic difference expansion (c-DE) technique, which leverages the 3D integer wavelet transform to embed watermark bits into neighboring voxels within cubes at low-frequency coefficients.
The voxel differences within each cube are expanded to create embedding space, and a majority voting mechanism is employed during extraction to enhance reliability. The embedding process incurs low distortion and supports lossless removal, thereby preserving the integrity and diagnostic accuracy of medical volume data.
Through these two benefits, Vol-Mark enables both integrity verification and ownership verification.
Integrity verification is first performed, and ownership verification through hypothesis testing is further conducted to enhance reliability, particularly under
data tampering or watermark removal attacks.
Comprehensive experimental results show the effectiveness of the proposed method and its superior robustness against conventional, geometric, and hybrid attacks on medical volume data.
In particular, through multiple tasks evaluations, Vol-Mark consistently achieves an ACC above 0.90 in most attack scenarios, outperforming existing methods by a clear margin.
\end{abstract}


\section{Introduction}
\begin{figure}[!htbp]
    \centering
    \includegraphics[width=0.9\linewidth]{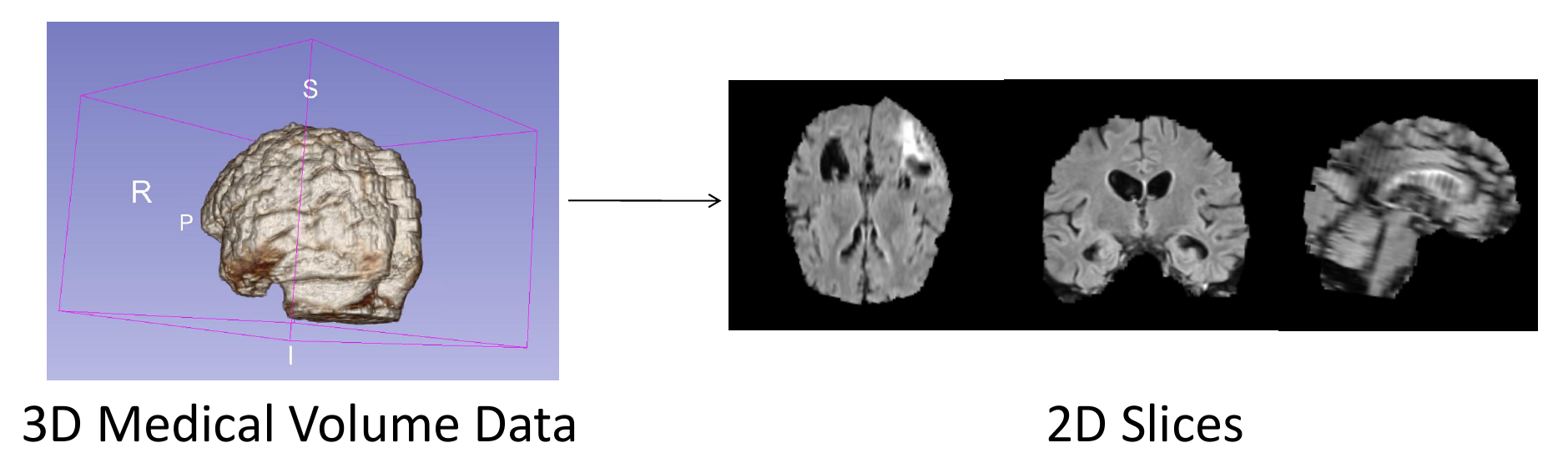}
    \caption{Medical volume data.}
    \label{fig:data}
\end{figure}

\IEEEPARstart{T}{he} rapid advancement of medical technologies, such as computed tomography (CT) and magnetic resonance imaging (MRI), has revolutionized medical diagnostics, telemedicine, and research~\cite{panayides2020ai,10620642,10258419, 10038494}. In recent years, there has been a growing shift towards representing medical data as 3D volume data (see Fig. \ref{fig:data}), which consists of multiple 2D slices. This approach provides a more comprehensive and detailed perspective, significantly enhancing diagnostic accuracy and enabling advanced medical research. However, the transmission and storage of 3D medical volume data in telemedicine pose significant challenges, particularly due to risks such as unauthorized access, privacy breaches, and data integrity concerns~\cite{anand2022hybrid, yan2022multiwatermarking, qureshi2026comprehensive}.

Watermarking is a widely used technology that enables healthcare institutions to securely store and transmit medical volume data while protecting it against unauthorized access and cyberattacks~\cite{wang2022image}.
Current watermarking methods for medical images can be broadly categorized
into three types: region-of-interest (ROI) lossless watermarking,
reversible watermarking, and zero-watermarking \cite{survey1}. ROI lossless watermarking preserves the diagnostic region by restricting embedding to the region of non-interest (RONI). However, the irreversible distortion introduced into the RONI can still compromise overall image fidelity and affect diagnostic accuracy \cite{ravichandran2021roi}. 

\IEEEpubidadjcol
Both zero-watermarking and reversible watermarking preserve data accuracy.
Zero-watermarking generates watermark information from extracted features without modifying the original data \cite{zero-concept}, making the
extraction of robust and discriminative features the key to effective zero-watermarking.
Contrastive learning~\cite{chen2020simple} was introduced to enhance
feature discriminability and watermarking robustness for medical data \cite{liu2025attack}.
However, these methods are still designed
around 2D images or average slices,  which limits their ability to capture volumetric features and reduces robustness against 3D specific attacks such as out-of-plane rotations.

Reversible watermarking ensures complete data recovery after watermark extraction, which have been extensively studied for 2D medical
images \cite{rever:arsalan2017protection, rever:wahed2019high, rever:gao2023efficient}. However, these approaches are specifically designed
for 2D medical images. Unlike images, medical 3D volume data are represented by voxels that exhibit spatial dependencies both within each slice and across adjacent slices, making direct extension of existing reversible methods challenging. To the best of our knowledge, no existing work has specifically studied reversible watermarking for medical volume data.

\begin{figure*}[!htbp]
\centerline{\includegraphics[width=1\textwidth]
{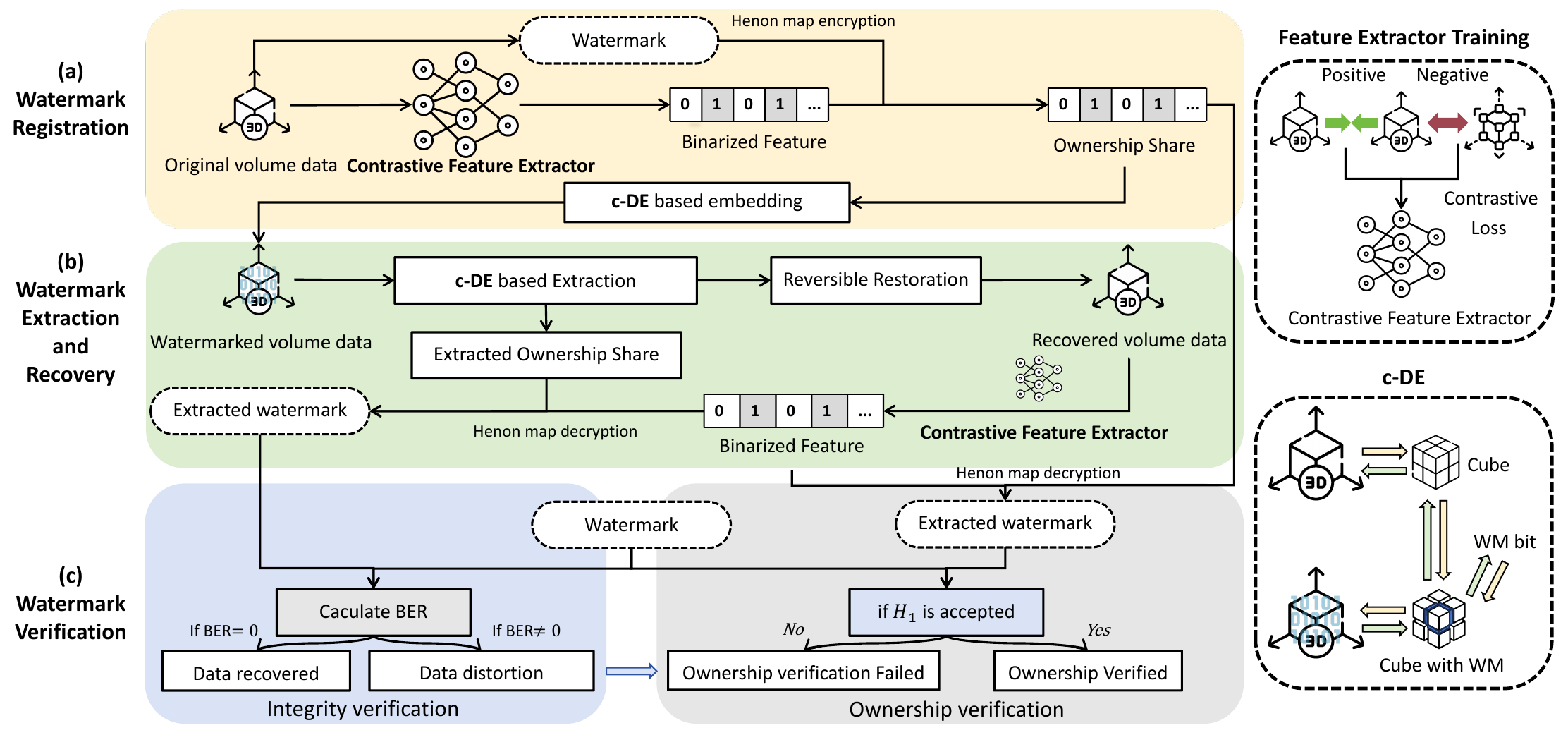}}

\captionsetup{justification=justified,singlelinecheck=false}
\caption{
    The workflow of our proposed Vol-mark method.
    (a) First, Vol-Mark extracts features from volume data using a designed contrastive feature extractor. With the encrypted watermark, Vol-Mark generates an ownership share, and embeds it into the data via the proposed c-DE technique.
    (b) Vol-Mark extracts the ownership share and restores the original data via the inverse c-DE, then retrieves the watermark from the extracted ownership share and features derived from the restored data. 
    (c) Vol-Mark applies double verification: integrity verification using $\text{BER}$ and ownership verification via hypothesis testing.
    }
\label{fig:Overview}
\end{figure*}

To address these limitations,
we propose a novel reversible-zero watermarking approach tailored for 3D medical volume data, termed Vol-Mark (see Fig.~\ref{fig:Overview}). Vol-Mark features the advantages of both reversible and zero watermarking,
offering lossless embedding and double verification capabilities.
Specifically, Vol-Mark leverages a contrastive loss to train a 3D ResNet-18 as feature extractor, enabling the extraction of robust volumetric features directly from volume data and  enhancing the discriminability and stability of the extracted features.
Furthermore, we introduce a novel technique, cubic difference expansion (c-DE), which is designed to embed watermark bit into cubes at low-frequency coefficients, ensuring reversible embedding and enhance reliability.
By integrating contrastive feature extraction with c-DE based reversible embedding, Vol-Mark achieves reliable watermark generation while preserving the integrity of the original data.

To summarize, the contributions of this paper are as follows.

(i) We introduce Vol-Mark, a novel reversible-zero watermarking approach designed for volume data.
Vol-Mark generates the zero watermark from features extracted via a novel volume feature extractor, which is designed to effectively extract discriminative and robust volumetric features via contrastive learning. Combined with the encrypted watermark, Vol-Mark ensures secure and reliable watermark generation.

(ii) Vol-Mark provides a novel c-DE technique that enables watermark reversibly embedding.
It embeds watermark bits into medical volume data via expanding voxel differences between neighboring voxels within and across slices into cubes at low-frequency coefficients, with the original voxel and watermark bits can be precisely recovered by inverting the expansion.
By utilizing c-DE, Vol-Mark does not require additional storage and allows the original data to be fully restored after watermark removal.

(iii) The proposed Vol-Mark provides both integrity verification and ownership verification.
Integrity verification is conducted first to detect potential tampering, and ownership verification is applied to improve the reliability of verification, even under data tampering or watermark removal attacks.

(iv) The proposed Vol-Mark demonstrates superior robustness against conventional, geometric, and their hybrid attacks in medical volume data. This indicates its capability to secure medical volume data in telemedicine applications, ensuring both ownership protection and data authenticity.

The rest of this paper is organized as follows. Section \ref{sec:related_work} reviews the related work. Section \ref{sec:preliminary} presents the preliminaries. Section \ref{sec:our_method} introduces the proposed method, and Section \ref{sec:experiments} presents the experimental results. Section \ref{sec:ablation} provides ablation studies. Finally, Section VII concludes the paper.

\section{Related Work}
\label{sec:related_work}

Conventional watermarking techniques typically embed watermarks in spatial or frequency domains~\cite{tao2014robust}.
In recent years, deep learning has been widely adopted in medical data for tasks such as segmentation~\cite{seg1, seg2, seg3}, diagnosis~\cite{diag1, diag2, diag3}, classification~\cite{class1} and image compression~\cite{compression}. The growing deployment of 
such models has also spurred research into model watermarking for ownership protection e.g. \cite{modelwm1, modelwm2, llmwm1}. For image watermarking, learning-based approaches have similarly been proposed e.g.~\cite{padhi2024deep, huang2023arwgan, internetwmrb1}. 
However, such techniques inevitably modify the original data, which is undesirable in diagnostic and medical research applications. 
Reversible watermarking and zero-watermarking are two types of watermarking methods that preserve data accuracy, 
as reviewed in detail below.

\begin{table*}[h]
\centering
\caption{A summary of existing reversible and zero watermarking methods for medical data.\\ $\checkmark$: supported; $\triangle$: partially supported; $\times$: not supported.}
\label{tab:related_work}
\renewcommand{\arraystretch}{1.3}
\begin{tabular}{p{3cm}p{4.9cm}cccp{3cm}}
\toprule
\textbf{Category} & \textbf{Representative approach} & \textbf{Reversible} & \textbf{Zero-WM} & \textbf{3D robustness} & \textbf{Limitation} \\
\midrule

Reversible WM (2D) \newline ~\cite{rever:wahed2019high,rever:arsalan2017protection,rever:gao2023efficient,bouarroudj2024high}
& Frequency-domain transforms& \checkmark & $\times$ & $\times$
& 2D only; no continuous ownership verification \\

Reversible-zero WM (2D) \newline 
~\cite{rovcek2016new,rzero1}
& Deep learning features + frequency-domain reversible embedding
& \checkmark & \checkmark & $\times$
& 2D only; no inter-slice 3D modeling \\
\midrule
Zero WM (2D) \newline
~\cite{zero-concept,han2023application,xiang2025trusted}
& Frequency-domain transforms or deep learning features
& $\times$ & \checkmark & $\times$
& Slice-based; limited feature distinguishability \\

Zero WM (3D volume) \newline
~\cite{3dzero3,3dzero1,LIU2021116124,liu2025attack}
& 3D frequency-domain transforms or slice-level deep features
& $\times$ & \checkmark & $\triangle$
& Vulnerable to out-of-plane transform \\

\midrule
Vol-Mark (ours)
& Volumetric deep features + cubic difference expansion based reversible embedding
& \checkmark & \checkmark & \checkmark
& — \\

\bottomrule
\end{tabular}
\end{table*}

\subsection{Zero Watermarking}

Zero-watermarking was first studied in~\cite{zero-concept}
by extracting stable image features using discrete cosine transform (DCT) and higher-order cumulants, without modifying the original data.
Following this, \cite{han2023application} proposed a federated learning-based scheme that trains a sparse autoencoder network to extract features, enabling zero-watermark generation while preserving patient privacy. \cite{xiang2025trusted} leveraged deep convolutional neural network (DCNN)-derived Gram matrices for stable feature extraction, combined with a hyperchaotic encryption system for enhanced security. While the above methods are designed for 2D images,
\cite{3dzero3} integrates 3D discrete wavelet transform (DWT), 3D discrete fourier transform, and a Hermite chaotic neural network for blind watermark extraction, while
\cite{3dzero1} employs 3D hyperchaos and 3D dual-tree complex wavelet transform (DTCWT) to extract low-frequency features.
However, medical images 
of the same organ share highly similar visual structures, which may limit the distinguishability 
of features. To address this, \cite{LIU2021116124} introduced ring statistics and an intra-slice variation mechanism to improve both robustness and distinguishability.
\cite{liu2025attack} proposed a contrastive learning-based framework 
that trains a dual-stream Siamese network to learn robust and discriminative features, achieving 
stronger resistance to both signal and geometric attacks.

\subsection{Reversible Watermarking}
Numerous reversible watermarking methods have been developed for 2D medical images.
For instance, \cite{rever:wahed2019high} introduced an interpolation-based reversible data hiding scheme that employs a capacity control parameter to determine the minimum embeddable bits per pixel.
\cite{rever:arsalan2017protection} embeds watermarks in the integer wavelet transform (IWT) domain via histogram shifting, leveraging genetic programming to achieve a better trade-off between imperceptibility and capacity. 
More recently, \cite{rever:gao2023efficient} combined Zernike moments with IWT to enhance robustness while maintaining reversibility. \cite{bouarroudj2024high} proposed a reversible fragile watermarking scheme for medical images that embeds an watermark encrypted via the chaotic Chen system into Discrete Fourier Transform frequency coefficients, achieving high embedding capacity and perfect reversibility.

Beyond purely reversible approaches, some methods integrate reversible and zero-watermarking to combine their respective strengths.
\cite{rovcek2016new} divides the image
into ROI and RONI, applying dual-tree complex wavelet transform-based zero-watermarking in the ROI and reversible contrast mapping in the RONI, where the secret share generated from the ROI is embedded into the RONI as a reversible watermark.
\cite{rzero1} employs VGG19-based feature extraction to construct an ownership share, which is then reversibly embedded using a combination of discrete wavelet transform, integer wavelet transform, and difference expansion, eliminating the need for third-party storage during verification.

\subsection{Watermarking for Medical Volume Data}

Unlike 2D images, watermarking for medical volume data remains relatively under-explored. 
Although 2D watermarking techniques can directly apply to slices of 3D volume data, such approaches
risk losing the spatial relationships inherent to the volumetric structure and may 
fail to accommodate the fundamental differences between pixel and voxel representations, limiting their robustness and applicability to medical volume data. Among existing watermarking methods for volume data, the predominant focus has been on 
zero-watermarking, where some approaches still  extract slice features \cite{LIU2021116124, liu2025attack}, failing to capture the inter-slice spatial structure. Such slice-based strategies offer limited robustness against transformations specific to 3D data, such as out-of-plane 
rotations. Furthermore, reversible watermark embedding for voxel-based data has yet to be 
investigated.
Compared with existing methods, our proposed Vol-Mark embeds the watermark reversibly by considering adjacent voxels and the spatial structure across different slices using the newly proposed c-DE algorithm, thereby preserving the continuity of 3D data.
Furthermore, a 3D ResNet-18 feature extractor based on contrastive learning is employed to extract high-level features for zero-watermark generation, which enhances both robustness and reliability.

\section{Preliminaries}
\label{sec:preliminary}
\subsection{Medical Volume Data}
Medical volume data is often represented as a real-valued three-dimensional array and described as
\begin{multline}
V = \left[v(i, j, k)\right]_{1 \leq i \leq M, 
1 \leq j \leq N,1 \leq k \leq O } \in \mathbb{R}^{M\times N\times O}
\end{multline}
where \( M \), \( N \) and \( O \) represent the dimensions of the three axes. Every $v(i, j, k)$  corresponds to a voxel (analogous to a pixel in 2D) and represents the intensity value in medical volume data, such as the attenuation coefficient (in CT) or the signal intensity (in MRI) at a specific location.

\subsection{Henon Map Encryption}
Henon map encryption \cite{Henon} 
is a chaotic dynamical system characterized by sensitive dependence on initial conditions,
and defined by the following recurrence equations
\begin{equation}
\begin{cases}
x_{n+1} = 1 - a x_n^2 + y_n, \\
y_{n+1} = b x_n
\end{cases}
\label{eq-henon}
\end{equation}
where $a$ and $b$ are system parameters that control the degree of chaos, commonly set to  $a=1.4$ and $b=0.3$ to ensure chaotic behavior. An initial condition for $(x, y)$, denoted as $(x_0, y_0)$, must be set to start the iterative process.

After performing the required number of iterations, the Henon map generates a chaotic sequence consisting of pairs $(x, y)$. 
To utilize this sequence for cryptographic applications, a thresholding technique
is often applied to transform the real-valued chaotic sequence into a binary chaotic sequence~\cite{threshold}. We use a simple thresholding technique~\cite{rzero1} to convert the chaotic sequence into binary. For a given number of bits $k$, the interval $[0,1)$ is uniformly partitioned 
into $2^k$ subintervals of equal length $\frac{1}{2^k}$. Each real number $x \in [0,1)$ is 
then mapped to an integer index
\begin{equation}
    i(x) = \left\lfloor 2^k \cdot x \right\rfloor, \quad i \in \{0, 1, \ldots, 2^k - 1\},
    \label{eq-thres}
\end{equation}
where $i(x) = i$ if and only if $x$ falls in the $i$-th interval, i.e.,
\begin{equation*}
    \frac{i}{2^k} \leq x < \frac{i+1}{2^k}.
\end{equation*}
The resulting index $i(x)$ is then encoded as its $k$-bit binary representation 
$b_{k-1} b_{k-2} \cdots b_0$, where $i = \sum_{j=0}^{k-1} b_j \cdot 2^j$, 
thereby mapping each real number $x \in [0, 1)$ to a $k$-bit binary sequence.

One key advantage of employing the Henon map for watermark scrambling is its extreme sensitivity to initial conditions, which ensures that the generated chaotic sequence is unique and unpredictable. This transforms the binary watermark into a highly randomized representation that cannot be recovered without the correct key, effectively preventing unauthorized forgery and enhancing the security of the proposed zero-watermarking scheme~\cite{zhang2023robust}.

\subsection{3D Integer Wavelet Transform}

The integer wavelet transform (IWT)~\cite{iwt} decomposes a signal or image into low and high frequency components. It is specifically designed to map integers to integers, making it highly suitable for lossless processing of discrete data. The inverse transform of IWT, denoted as IWT$^{-1}$, 
can precisely reconstruct the original signal from the components generated by the IWT. By combining both IWT and IWT$^{-1}$, a fully reversible transformation is achieved. IWT often employs lifting schemes, which enable precise reconstruction of the original signal without introducing numerical errors, even in limited-precision environments. 

\begin{figure}[!htbp]
    \centering
    \includegraphics[width=1\linewidth]{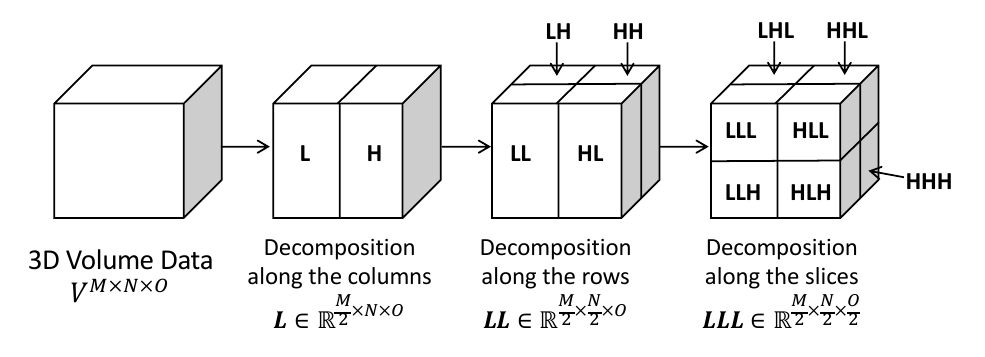}
    \caption{3D integer wavelet transform scheme.}
    \label{fig:3diwt}
\end{figure}

The 3D-IWT decomposes 3D volume data into representative volumetric features. Specifically, as shown in Fig.~\ref{fig:3diwt}, after applying a single-level 3D-IWT, the transformation produces low-frequency component coefficients (LLL), which retain the most significant structural information.
It also generates seven sets of high-frequency detail components (LLH, LHL, LHH, HLL, HLH, HHL, and HHH)~\cite{3diwtfig}, which capture finer variations in the data. By utilizing the inverse transform, 3D-IWT$^{-1}$, the coefficients generated by 3D-IWT can be precisely reconstructed back into the original volume data. This precise reversibility makes 3D-IWT a ideal tool for reversible watermarking, where the original volume data must be exactly recovered after watermark extraction.

\section{Proposed Method}
\label{sec:our_method}
In this section, we present our Vol-Mark method.
It consists of three phases: watermark registration, extraction and recovery, and verification. 

\subsection{Watermark Registration}
\label{subsec:wmregis}
Watermark registration phase consists of three processes: feature extraction, ownership share generation and embedding. 

\subsubsection{Feature extraction for volume data}
\label{subsubsec:fextra}
Existing transform-based methods offer limited geometric stability, while approaches 
relying on images or 2D slice averaging neglect inter-slice spatial correlations. 
To address this, we design a contrastive learning-based feature extractor using 3D ResNet-18, inspired by \cite{liu2025attack}, to capture discriminative and robust volumetric features from the entire volume data.

\begin{figure}[!htbp]
    \centering
    \includegraphics[width=1\linewidth]{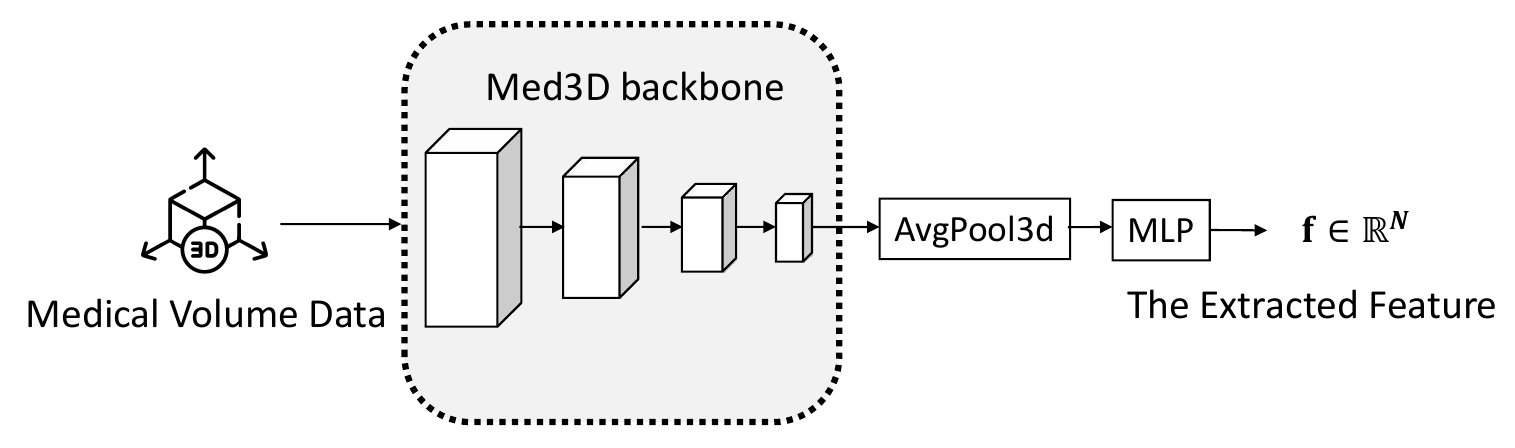}
    \caption{Architecture of the feature extractor.}
    \label{fig:3dmodel}
\end{figure}

\textit{(i) Network architecture.}
To leverage prior knowledge from large-scale medical datasets, we adopt the pretrained 3D ResNet-18 model from MedicalNet \cite{chen2019med3d} as the backbone, 
whose 3D convolutional kernels effectively capture inter-slice spatial correlations 
within volume data \cite{wang2022medical}. The classification head is removed, 
and a three-layer multi-layer perceptron (MLP) projector with batch normalization is then applied to map the backbone output into a feature vector $\mathbf{f} \in \mathbb{R}^N$, where $N$ denotes the watermark bit length.
The overall architecture of the feature extractor is illustrated in Fig.~\ref{fig:3dmodel}.

\textit{(ii) Preprocessing.}
All input volumes are resized to the same
resolution of $128 \times 128 \times 64$. The data are then normalized using the global mean $\mu$ and standard deviation $\sigma$ computed from the training dataset via $V_{norm} = \frac{V - \mu}{\sigma}$.

Data augmentation is applied to the original medical volumes to generate positive
pairs for contrastive learning. Specifically, each volume is randomly subjected
to one or more of the following transformations: Gaussian noise with variance
in $[0.01, 0.25]$, salt-and-pepper noise with density in $[0.01, 0.15]$,
JPEG compression with quality factor in $[50, 90]$, median or average filtering
with kernel size selected from $\{3, 5, 7\}$, cropping with ratio in $[0.01, 0.20]$,
rotation within $[1^\circ, 30^\circ]$, translation with magnitude in $[0.01, 0.20]$
along a randomly selected axis or plane, and random dropping with drop ratio
in $[0.01, 0.25]$.

\textit{(iii) Loss function.}
To ensure the extracted features are both distinctive and stable across different instances of the same image, we adopt a contrastive loss \cite{chen2020simple}
formulated as
\begin{equation}
\mathcal{L}_{con} =
- \log 
\frac{\exp(\text{sim}(z_i^{(1)}, z_i^{(2)})/\tau)}
{\sum_{k \neq i} \exp(\text{sim}(z_i, z_k)/\tau)}
\end{equation}
where $z_i^{(1)}$ and $z_i^{(2)}$ denote the features of two augmented views of the $i$-th sample, forming a positive pair. All other samples $z_k$ with index $k\neq i$ in the batch are treated as negative samples.
$\text{sim}(\cdot)$ represents cosine similarity, and $\tau$ is a temperature parameter.

\textit{(iv) Feature binarization.}
To obtain binary features, we apply a threshold of zero to the extracted feature vector $\mathbf{f} = (f_1, f_2, \ldots, f_N)$, where zero serves 
as a natural threshold owing to the batch normalization applied in the projector. 
The binary feature vector $\mathbf{f^b} = (f^b_1, f^b_2, \ldots, f^b_n)$ 
is obtained as

\begin{equation}
f^b_i =
\begin{cases}
1, & f_i > 0, \\
0, & \text{otherwise}.
\end{cases}
\label{eq:binarization}
\end{equation}

By training a 3D ResNet-18 feature extractor with contrastive loss, Vol-Mark successfully learns distinctive and robust binarized features for watermark generation.

\subsubsection{Ownership share generation}
\label{subsubsec:owngen}
The ownership share $\mathbf{OS}$ is generated by combining the preset watermark 
$\boldsymbol{w}$, a binary chaotic sequence $\mathbf{c^b}$, and the binary feature 
vector $\mathbf{f^b}$ via XOR operation.
Specifically, the initial parameters of the Henon map are set, optionally derived 
from patient information or other metadata, and iterated via \eqref{eq-henon} to generate a chaotic sequence $\mathbf{c}$, which is then binarized into $\mathbf{c^b}$ using \eqref{eq-thres}.
An XOR operation is performed among the preset watermark $\boldsymbol{w}$, $\mathbf{c^b}$ and $\mathbf{f^b}$ to produce  the ownership share $\mathbf{OS}$:
\[
\mathbf{OS} = \boldsymbol{w} \oplus \mathbf{c^b} \oplus \mathbf{f^b} ,
\]
which serves as the watermark for embedding and extraction in the subsequent process.

\subsubsection{Ownership share embedding}

\begin{figure}[!htbp]
    \centering
    \includegraphics[width=1\linewidth]{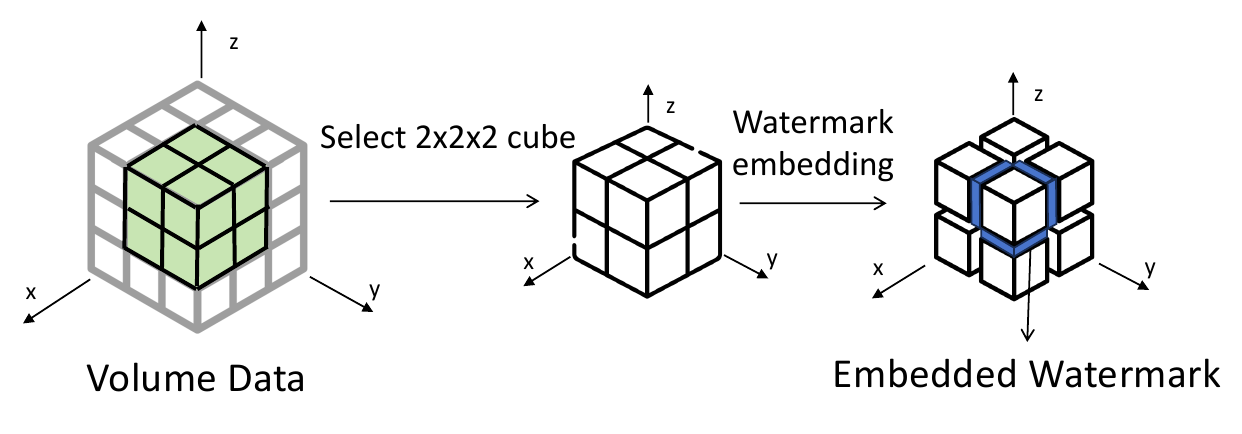}
    \caption{Cubic difference expansion (c-DE)}
    \label{fig:c-DE}
\end{figure}

We propose a new \textit{Cubic Difference Expansion} (c-DE) method
for reversible embedding in volume data (see Fig. \ref{fig:c-DE})\footnote{An early version of part of this work was presented in
\cite{zhu2025reversible}.}.
Unlike 2D embedding approaches~\cite{DE}, which operate solely on individual images and exploit only horizontal adjacency, Vol-Mark captures voxel relationships in 3D volumes by considering both horizontal and vertical adjacency. This extension to volume data enables a more comprehensive use of inter-voxel dependencies, where differences between neighboring voxel values are leveraged and expanded (typically by doubling) to create space for embedding verification information 
such as a watermark or ownership share.
In particular, c-DE divide the volume data into $2 \times 2 \times 2$ cubes and embed a binary bit into each cube. 
For every cube, the bits are embedded into the differences between a reference point and its three neighboring points (see Fig. \ref{fig:c-DE}).
In addition, c-DE ensures reversibility, as the embedding process is fully based on integer operations.
It employs 3D-IWT, whose reversibility guarantees accurate data
extraction without any loss. All other operations in c-DE also integer-based, which further ensures lossless reconstruction and preserves the reversibility of the overall embedding process.

The workflow of c-DE for embedding $\mathbf{OS}$ into volume data is summarized in Algorithm~\ref{alg:cde-embed}. The detailed steps are as follows:

\textit{Step 1. 3D-IWT transform.} 
Apply a 3D-IWT transform to the medical volume data $V_{ori}$, producing eight sub-bands:
\begin{align*}
    \{LLL_{int}, HLL_{int}, \dots, HHH_{int} \} \leftarrow \text{3D-IWT}(V_{ori})
\end{align*} 
The low-frequency $LLL$ sub-band is used for embedding, which contains the most significant structural information of the volume data and ensures the stable of embedded $\mathbf{OS}$.

\textit{Step 2. Divide cubes. }  The $LLL$ sub-band is divided to $2\times2\times2$ cubes. 
Denote each cube as \[
S = \{s(i,j,k) \mid i, j, k \in \{0, 1\}\}
\]
where $i, j, k$ are the coordinates along the three axes $x, y, z$. We use 4 points: a reference point $A$ and its three neighboring points $B, C, D$ to embed $\mathbf{OS}$ bits.

\textit{Step 3. Calculate the difference. }
Calculate the difference between $A$ and its selected neighbors $B, C, D$ as 
\begin{align}
\label{eq:cal_difference}
    d_{ab} = A-B, \quad d_{ac} = A -C, \quad d_{ad} = A - D
\end{align}
and their average as
$M = (A + B + C + D)/{4}.$

\textit{Step 4. Expand difference and embed ownership share.}
 Expand the differences by doubling their original value, $d' = 2d$, to create space for embedding. Next, embed the $\mathbf{OS}$ bit into the expanded differences by $d'' = d' + b$. 

\textit{Step 5. Rebuild voxel values.}
Use the watermarked differences $d''$ and their average to rebuild new voxel values, effectively incorporating the $\mathbf{OS}$ bit.
\begin{align}
\label{eq:reconstruct_voxel}
    \nonumber&A' = \text{Round}_{\downarrow}( M - (d''_{ab} + d''_{ac} + d''_{ad})/{4} ),\\
    B' &= A' - d''_{ab}, \quad
C' = A' - d''_{ac}, \quad
D' = A' - d''_{ad}
\end{align}
where 
$\text{Round}_{\downarrow}\left( x \right) = \lceil x - 0.5 \rceil.$

\textit{Step 6. Overflow/underflow check.}
Check whether the rebuilt voxels exceed the valid range, and only renew the cubes that pass the overflow check to get $LLL_{wm}$. If overflow occurs, mark the current cube in the location map $L$.

\textit{Step 7. Rebuild volume data.} Rebuild watermarked medical volume data $V_{wm}$ through inverse 3D-IWT transform.
\[
V_{wm} \leftarrow \text{3D-IWT}^{-1}(\{LLL_{wm}, HLL_{int}, \dots, HHH_{int} \})
\]

\begin{algorithm}[t]
\caption{Cubic Difference Expansion (c-DE)}
\label{alg:cde-embed}
\begin{algorithmic}[1]
\State \textbf{Input}: Ownership share $\mathbf{OS}$, original volume data $V_{ori}$
\State \textbf{Output}: Watermarked volume data $V_{wm}$, location map $L$
\State Initialize location map $L$.
\State $\{LLL_{int}, HLL_{int}, \dots, HHH_{int} \} \leftarrow \text{3D-IWT}(V_{ori}) $
\For{each $2\times2\times2$ cube in $LLL_{int}$}
    \State $(A, B, C, D)\leftarrow$ SelectPoints(cube)
    \State
    $(d_{ab}, d_{ac}, d_{ad})\leftarrow$ 
    CalculaDiff$(A, B, C, D)$ via \eqref{eq:cal_difference}
    \State Embed watermark bit: 
        \[
        \begin{aligned}
        d''_{ab} &= 2 \cdot d_{ab} + \mathbf{OS}[next\_bit], \quad \\ d''_{ac} &= 2 \cdot d_{ac} + \mathbf{OS}[next\_bit], \quad \\
        d''_{ad} &= 2 \cdot d_{ad} + \mathbf{OS}[next\_bit]
        \end{aligned}
        \]
    \State Rebuild voxel values $(A', B', C', D')$ via \eqref{eq:reconstruct_voxel}
    \If{any of $A', B', C', D'$ exceeds the valid range}
        \State Mark current cube in $L$.
    \Else
        \State $(A', B', C', D') \rightarrow$ Renew(cube)
    \EndIf
\EndFor
\State $LLL_{wm} \leftarrow$ updated $LLL_{int}$
\State $V_{wm} \leftarrow \text{3D-IWT}^{-1}(\{LLL_{wm}, HLL_{int},  \dots, HHH_{int} \})$
\State \textbf{Return} $V_{wm}, L$.
\end{algorithmic}
\end{algorithm}

Through watermark registration, Vol-Mark generates the ownership share $\mathbf{OS}$ and embeds it into the original volume data via the c-DE algorithm. $\mathbf{OS}$ is produced by combining unique features extracted from the volume with the preset watermark $\boldsymbol{w}$, followed by encryption. This process tightly couples the watermark with the data content, achieving low-distortion embedding while simultaneously registering a zero-watermark to guard against unauthorized removal or forgery.

\subsection{Watermark Extraction and Recovery}
In the watermark extraction and recovery phase, Vol-Mark extracts the watermark from watermarked data and restores the original data. 

\subsubsection{Ownership share extraction and data recovery}
The inverse of c-DE is applied to recover the ownership share $\hat{\mathbf{OS}}$ from the watermarked data and reconstruct  the original volume data.
The ownership share is extracted by analyzing the difference values of the three neighboring points. Ideally, the extracted bit of ownership share should be identical across all three points. To address discrepancies, we introduce a majority voting mechanism: the extracted bit is set to 1 if at least two of the three bits are 1; otherwise, it is set to 0.
This enhances the reliability of ownership share extraction.

A summary of the inverse c-DE process is provided in Algorithm~\ref{alg:cde-extract}, which  consists of the following steps.

\begin{algorithm}[t]
\caption{Inverse c-DE}
\label{alg:cde-extract}
\begin{algorithmic}[1]
\State \textbf{Input}: Watermarked volume data $V_{wm}$, location map $L$
\State \textbf{Output}: Extracted ownership share $\hat{\mathbf{OS}}$, original volume data $V_{ori}$.
\State Initialize extracted watermark $\hat{\mathbf{OS}} \leftarrow \{\}$.
\State $\{LLL_{wm}, HLL_{int}, \dots, HHH_{int} \} \leftarrow \text{3D-IWT}(V_{wm})$
\For{each $2\times2\times2$ cube in $LLL_{wm}$}
    \If{cube not in $L$}
        \State $(A', B', C', D')\leftarrow$ SelectPoints(cube)
        \State
        $(d''_{ab}, d''_{ac}, d''_{ad})\leftarrow$ 
    CalculaDiff$(A, B, C, D)$ via \eqref{eq:cal_difference_inverse}
        \If{$\vert$\{$X$\text{ is odd}\,$: X \in \{d''_{ab}, d''_{ac}, d''_{ad}$\}$\vert$$\ge 2$}
        \State $b\leftarrow 1$
        \Else 
        \State $b\leftarrow 0$
        \EndIf
        \State Append $b$ to $\hat{\mathbf{OS}}$
        \State Rebuild voxel values of $(A, B, C, D)$ via \eqref{eq:rebuild_reverse}
        \State $(A, B, C, D) \rightarrow$ Renew(cube)
    \EndIf
\EndFor
\State $LLL_{ori} \leftarrow$ updated $LLL_{wm}$
\State $V_{ori} \leftarrow \text{3D-IWT}^{-1}(\{LLL_{ori}, HLL_{int}, \dots, HHH_{int} \})$
\State \textbf{Return} $\hat{\mathbf{OS}}, V_{ori}$.
\end{algorithmic}
\end{algorithm}

\textit{Step 1. 3D-IWT transform.}
Apply a 3D-IWT transform to the watermarked volume data $V_{wm}$, producing eight sub-bands, and obtain the $LLL_{wm}$.
\begin{align*}
    \{LLL_{wm}, HLL_{int}, \dots, HHH_{int} \} \leftarrow \text{3D-IWT}(V_{wm})
\end{align*}

\textit{Step 2. Divide cubes. } 
Divide the $LLL_{wm}$ into $2 \times 2\times 2$ cubes and obtain $A', B', C', D'$  using the same method as in the c-DE embedding process. Skip cubes marked in $L$.

\textit{Step 3. Calculate difference.} Calculate the expanded differences 
\begin{align}
\label{eq:cal_difference_inverse}
    d''_{ab} =  A' - B', \ 
d''_{ac} = A' - C',\ 
d''_{ad} = A' - D' 
\end{align}
and their average as
$M = (A' + B' + C' + D')/{4}.$

\textit{Step 4. Remove ownership and restore differences.} Remove the embedded $\mathbf{OS}$ bit $b$, then halve and restore the original difference values via
\begin{align*}
    d_{ab} =  (d''_{ab} - b)/2, \
d_{ac} = (d''_{ac} - b)/2,\ 
d_{ad} = (d''_{ad} - b)/2. 
\end{align*}

\textit{Step 5. Restore voxel values.}
Use the above restored differences and their average to restore original voxel values $(A, B, C, D)$.
\begin{align}
\label{eq:rebuild_reverse}
\nonumber&A =  \text{Round}_{\uparrow}(M - (d_{ab} + d_{ac} + d_{ad}) / {4}),\\
B &= A - d_{ab}, \quad C = A - d_{ac}, \quad D = A - d_{ad}
\end{align}
where
$\text{Round}_{\uparrow}\left( x \right) =  \lfloor x + 0.5 \rfloor.$

\textit{Step 6. Restore voxels.}
Renew the $(A, B, C, D)$ of the cubes to obtain the original sub-band $LLL_{ori}$.

\textit{Step 7. Restore volume data.} Rebuild original medical volume data $V_{ori}$ through inverse 3D-IWT transform.
\[
V_{ori} \leftarrow \text{3D-IWT}^{-1}(\{LLL_{ori}, HLL_{int}, \dots, HHH_{int} \})
\]

\subsubsection{Watermark Extraction}
\label{sec:wmextra}
The watermark is recovered as follows. Features $\hat{\mathbf{f}}$ are extracted from the restored volume as described in Section~\ref{subsubsec:fextra}, and the chaotic sequence is regenerated using the same initial value as in Section~\ref{subsubsec:owngen}. The extracted watermark $\hat{\boldsymbol{w}}$ is then obtained via XOR among the binarized features, the chaotic sequence, and $\hat{\mathbf{OS}}$:
\[
\hat{\boldsymbol{w}} = \hat{\mathbf{OS}} \oplus \mathbf{c}^{\mathbf{b}} \oplus \hat{\mathbf{f}}.
\]

Through watermark extraction and recovery, Vol-Mark applies the inverse of the c-DE algorithm to recover the original volume data $V_{ori}$ and extract the embedded ownership share $\hat{\mathbf{OS}}$. The extracted watermark $\hat{\boldsymbol{w}}$ is subsequently decoded by combining unique features and the same chaotic sequence, enabling reliable ownership verification while ensuring complete restoration of the original data.

\subsection{Watermark Verification}
Vol-Mark achieves double verification by using both reversible and zero watermarking schemes.

\subsubsection{Integrity verification}
The extracted watermark $\boldsymbol{\hat{w}}$
is extracted using the extracted $\hat{\mathbf{OS}}$ following \ref{sec:wmextra}.
It is compared with the stored original one, and the Bit Error Rate (BER) value is calculated via
\begin{equation}
\text{BER} \triangleq \frac{1}{N} \lVert \boldsymbol{w} - \hat{\boldsymbol{w}}\rVert_1
\label{eq:ber}
\end{equation}
where \( N \) represents the total number of watermark bits, $\boldsymbol{w}$ is the original watermark, and $\boldsymbol{\hat{w}}$ is the extracted watermark.

If BER is zero, it indicates that the original data is not altered and the source data can be losslessly recovered after watermark extraction. Conversely, if BER is not zero, it suggests that the data is subjected to distortion. In such cases, the zero-watermarking scheme is used instead. The pre-stored ownership share $\mathbf{OS}$ is utilized to verify the watermark.

\subsubsection{Ownership verification}
We utilize hypothesis testing to verify the extracted watermark.
We employ a \textit{binomial test} to determine whether the watermark is detected, defining the test as whether the
number of correctly detected bits significantly exceeds what would be expected randomly. The null hypothesis $H_0$ and alternative hypothesis $H_1$ are defined as:
\begin{align}
    H_0&: \xi = 0.5 \quad \text{(watermark not detected)}, \notag \\
    H_1&: \xi > 0.5 \quad \text{(watermark detected)}.
\end{align}

For a watermark of length $N$, let $k$ represent the number of successfully matched bits. Under $H_0$, the number of matched bits $X \sim \mathcal{B}(N, 0.5)$. The p-value is calculated using a right-tailed test:
\begin{equation}
    p = \text{Pr}[X \geq k \mid H_0] = \sum_{i=k}^{N} \binom{N}{i} (0.5)^{N}.
\end{equation}
We adopt the statistical criterion recommended in~\cite{COX200815} to ensure the reliability of the watermark verification, $\alpha = 10^{-6}$. We reject $H_0$ if $p \leq \alpha$.

To verify whether the selected significance level is appropriate, we conducted experiments using volumes from the Task01 of MSD dataset \cite{dataset}. One volume data was selected as the watermarked data, while the remaining 400 volumes were not. Watermark extraction was performed 400 times. The average p-values are shown in Table~\ref{tab:pvalue}, which clearly indicate the presence of the watermark. The p-values obtained from the watermarked data are clearly distinguishable from those of the other volumes, demonstrating strong discriminative capability. These results show that the selected significance level can effectively distinguish the watermarked data from non-watermarked data.

Through watermark verification, Vol-Mark performs double verification via both integrity and ownership checks. When $\text{BER} = 0$, the embedded watermark is directly recovered and ownership is confirmed. When $\text{BER} > 0$, the pre-stored $\mathbf{OS}$ is used for zero-watermark verification, ensuring robustness against data distortion.

\begin{table}
    \centering
    \caption{P-value results of watermarked data and others.}
    \resizebox{\linewidth}{!}{ 
    \begin{tabular}{ccc}
    \toprule
         & Watermarked data & Non-watermarked data\\
    \midrule
        p-value & $7.46\times10^{-155} \pm 0 $ & $0.5264 \pm 0.4535 $\\
        Result & Watermark detected & Watermark not detected \\
    \bottomrule
    \end{tabular}
    }
    \label{tab:pvalue}
\end{table}

\section{Evaluations}
\label{sec:experiments}

\begin{table*}[!htbp]
\caption{NC and BER results under conventional attacks.}
\begin{center}
\begin{tabular}{c|c|ccc|ccc|ccc}
\toprule
\multirow{2}{*}{\textbf{Types of attacks}} & \multirow{2}{*}{\textbf{Intensity}} 
  & \multicolumn{3}{c}{\textbf{Task01} (Brain Tumours)}& \multicolumn{3}{c}{\textbf{Task07} (Pancreas)}&  \multicolumn{3}{c}{Average}\\ 
\cline{3-11}
 & & PSNR$\uparrow$& BER$\downarrow$ &NC$\uparrow$& PSNR$\uparrow$& BER$\downarrow$  &NC$\uparrow$& PSNR$\uparrow$& BER$\downarrow$  &NC$\uparrow$\\ 
\midrule

\multirow{5}{*}{Gaussian noise} 
& 1\% & 37.08 & 0.0027 &0.9965 & 37.01 & 0.0022 &0.9972 & 37.05 & 0.0025 &0.9969 
\\ 
 & 5\% & 23.56 & 0.0127 &0.9839 & 23.85 & 0.0139 &0.9823 & 23.71 & 0.0133 &0.9831 
\\ 
 & 10\% & 17.55 & 0.0260 &0.9671 & 17.85 & 0.0286 &0.9640 & 17.70 & 0.0273 &0.9656 
\\ 
 & 20\% & 11.54 & 0.0597 &0.9254 & 
11.84 & 
0.0513 &0.9358 & 11.69 & 0.0555 &0.9306 
\\ 
 & 25\% & 9.60 & 0.0776 &0.9036 & 9.90 & 0.0611 &0.9237 & 9.75 & 0.0694 &0.9137 
\\ 
\midrule
\multirow{3}{*}{Salt-and-pepper noise} 
& 1\% & 22.33 & 0.0173 &0.9781 & 22.95 & 0.0132 &0.9833 & 22.64 & 0.0153 &0.9807 
\\ 
 & 3\% & 17.54 & 0.0322 &0.9594 & 18.18 & 0.0271 &0.9659 & 17.86 & 0.0297 &0.9627 
\\ 
 & 5\% & 15.35 & 0.0439 &0.9449 & 15.99 & 0.0386 &0.9514 & 15.67 & 0.0413 &0.9482 
\\ 

 & 10\% & 12.36 & 0.0702 &0.9127 & 13.00 & 0.0633 &0.9209 & 12.68 & 0.0668 &0.9168 
\\ 

 & 15\% & 10.57 & 0.0901 &0.8887 & 11.22 & 0.0894 &0.8894 & 10.90 & 0.0898 &0.8891 
\\ 

\midrule
\multirow{5}{*}{JPEG Compression} 
& 50\%& 37.96 & 0.0058 &0.9927 & 
33.91 & 
0.0089 &0.9887 & 35.94 & 0.0074 &0.9907 
\\ 
& 60\%& 38.64 & 0.0054 &0.9931 & 34.66 & 0.0082 &0.9896 & 36.65 & 0.0068 &0.9914 
\\ 
& 70\%& 39.47 & 0.0056 &0.9929 & 
35.69 & 
0.0074 &0.9906 & 37.58 & 0.0065 &0.9918 
\\ 
& 80\%& 40.96 & 0.0056 &0.9929 & 37.25 & 0.0080 &0.9898 & 39.11 & 0.0068 &0.9914 
\\ 
& 90\%& 43.53 & 0.0053 &0.9933 & 
39.97 & 
0.0082 &0.9896 & 41.75 & 0.0068 &0.9915 
\\ 
\midrule
\multirow{2}{*}{Median filtering} 
& 3 & 35.61 & 0.0174 &0.9780 & 30.40 & 0.0358 &0.9549 & 33.01 & 0.0266 &0.9665 
\\ 
& 5& 32.29 & 0.0405 &0.9492 & 27.94 & 0.0565 &0.9295 & 30.12 & 0.0485 &0.9394 
\\
& 7& 30.51 & 0.0633 &0.9215 & 26.99 & 0.0752 &0.9066 & 28.75 & 0.0693 &0.9141 
\\
\midrule
\multirow{2}{*}{Average filtering} 
& 3 & 33.55 & 0.0117 &0.9851 & 29.47 & 0.0156 &0.9802 & 31.51 & 0.0137 &0.9827 
\\ 
& 5& 30.56 & 0.0291 &0.9633 & 27.10 & 0.0359 &0.9549 & 28.83 & 0.0325 &0.9591 
\\
& 7& 28.95 & 0.0451 &0.9434 & 25.83 & 0.0530 &0.9338 & 27.39 & 0.0491 &0.9386 
\\

\bottomrule
\end{tabular}
\label{tab:ber1}
\end{center}
\end{table*}

\begin{figure}[htbp]
    \centering
    \subfloat[Task01 Brain Tumours]{
        \includegraphics[width=\linewidth]{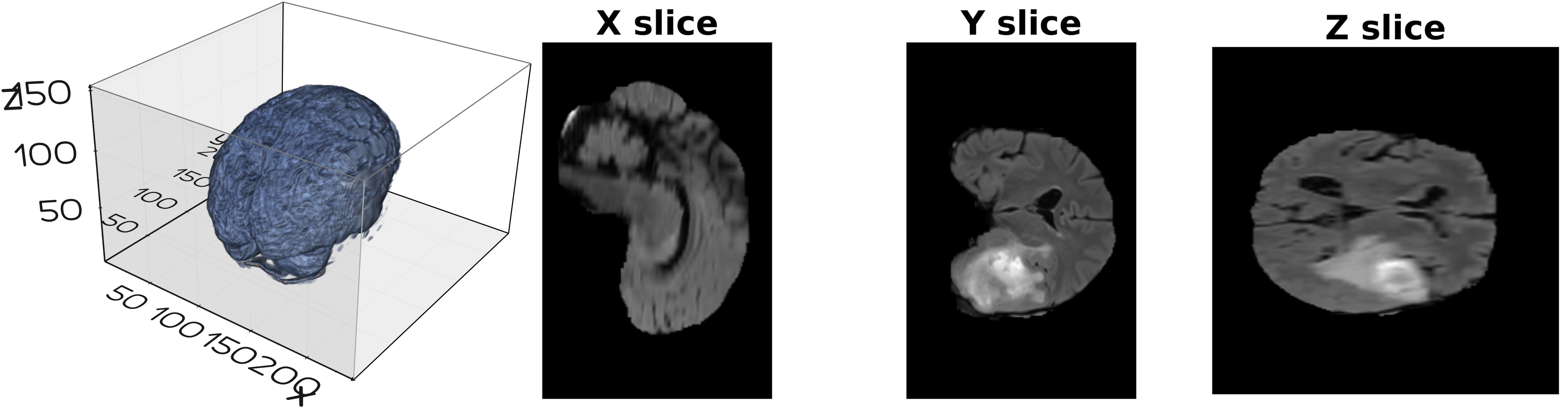}
        \label{fig:01}
    }
    \par\medskip
    \subfloat[Task03 Liver]{
        \includegraphics[width=\linewidth]{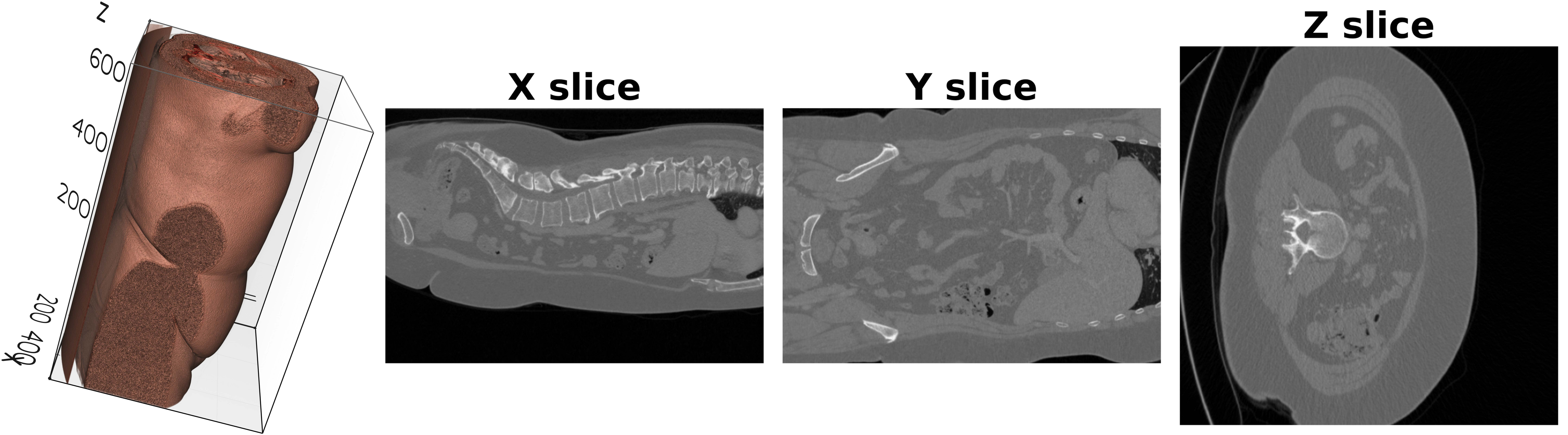}
        \label{fig:03}
    }
    \par\medskip
    \subfloat[Task07 Pancreas]{
        \includegraphics[width=\linewidth]{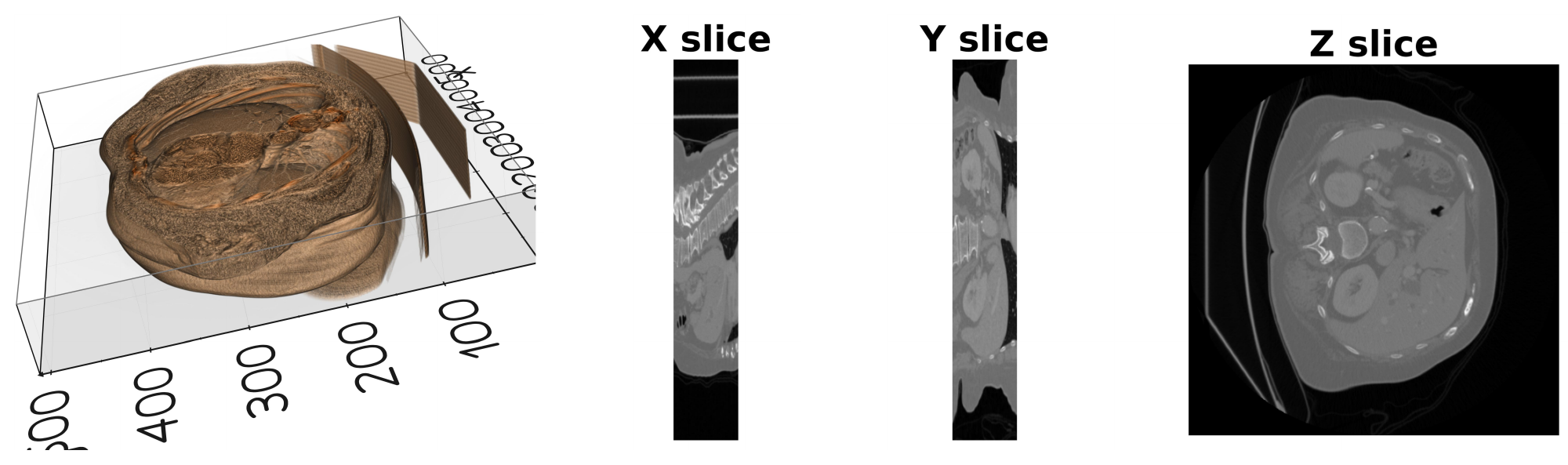}
        \label{fig:07}
    }
    \par\medskip
    \caption{The slices and 3D models of samples in MSD dataset.}
    \label{fig:mriimage}
\end{figure}

\textbf{Experimental setup.} 
We evaluate the performance of our method on MSD dataset \cite{dataset}, which is a comprehensive collection designed for medical semantic segmentation challenges, encompassing 10 different types of medical 3D/4D images.
As summarized in Table \ref{tab:msd}, 
three tasks are used in 
our experiments: Task01 (Brain Tumours) consists of 750 4D brain MRI scans, Task03 (Liver) contains 210 
3D CT scans with annotations of the liver and liver tumours, and Task07 (Pancreas) comprises 
420 3D CT volumes for the pancreas and pancreatic tumours. These three tasks are adopted as they offer sufficient training data of more than 200 volumes each, cover both MRI and CT modalities, and contain large and varied data shapes across tasks.
From each Task01 scan, the 3D volume at the first index of the fourth dimension is extracted for our experiments.
Fig.~\ref{fig:mriimage} shows the central slices along three axes and the corresponding 3D models of
samples from the MSD dataset.

\begin{table}[h]
    \centering
    \caption{Tasks in MSD dataset.}
    \begin{center}
    \resizebox{\linewidth}{!}{
        \begin{tabular}{ccccc}\toprule
            ID & Task &  Modality& Size & Median Shape\\\midrule
            Task01 & Brain Tumours &  MRI&  750 4D volumes& $240\times 240\times155\times4$\\
             Task03&  Liver&  CT& 210 3D volumes& $512\times512\times391$\\
             Task07&  Pancreas &  CT &  420 3D volumes& $512\times512\times93$\\ 
             \bottomrule
        \end{tabular}}
    \label{tab:msd}
    \end{center}
\end{table}

We randomly select 200 volumes from each task, with 150 volumes used for training and the remaining 50 volumes used for testing. The batch size is set to 16, and the model is trained with 200 epochs for each task. We adopt Adam optimizer with a weight decay of $1\times10^{-5}$. A cosine annealing learning rate scheduler is used, with an initial learning rate of $1\times10^{-4}$ and a minimum learning rate of $1\times10^{-6}$. The temperature parameter is set to 0.05.
For evaluation, we report the average results of each task on the test set.

\textbf{Evaluation metrics.}
We adopt three commonly-used metrics PSNR, BER and NC to evaluate
the performance of the proposed method. 
The PSNR value, defined as in \eqref{eq:psnr}, presents the degree to which the medical volume data has been distorted by various attacks.
The BER value as in \eqref{eq:ber} is used to evaluate the bit-wise error between the original and extracted watermark, while the NC value defined in \eqref{eq:nc} measures their overall similarity.
\begin{equation}
\text{PSNR} = 10 \cdot \log_{10} \left( \frac{\text{MAX}^2}{\sum \left( v - \hat{v} \right)^2} \right)
\label{eq:psnr}
\end{equation}
\begin{equation}
\text{NC} = \frac{\sum \boldsymbol{w} \cdot \boldsymbol{\hat{w}}}{\sqrt{\sum \boldsymbol{w}^2} \cdot \sqrt{\sum \boldsymbol{\hat{w}}^2}}    
\label{eq:nc}
\end{equation}
where $\text{MAX}$ denotes the maximum possible voxel intensity in the 3D volume data,
$v$ and $\hat{v}$ refer to the voxel in the original 3D volume data and in the restored volume data, respectively.

\begin{table}[!htbp]
\caption{PSNR and BER results without attacks}
\begin{center}
\resizebox{\linewidth}{!}{ \begin{tabular}{c|c|c|c}
\toprule
Task& PSNR$\uparrow$ (watermarked) & PSNR$\uparrow$ (recovered) & \textbf{BER}$\downarrow$ \\ 
\midrule
Task01
(Brain Tumours)& 95.51& / & 0.00 \\
 Task03
(Liver)& 82.10& / &0.00 \\
 Task07
(Pancreas)& 76.34& / &0.00 \\
\midrule
 Average& 84.65& / &0.00 \\
\bottomrule
\end{tabular}
}
\label{tab:ber0}
\end{center}
\end{table}

\subsection{Performance without Attacks}
We verify the reversibility of Vol-Mark in the absence of attacks.
As reported
in Table \ref{tab:ber0},
the recovered PSNR value is infinite and the BER is zero. This indicates that Vol-Mark achieves perfect reversibility, allowing for the original volume data to be fully recovered without any distortion or loss of information when no attacks are applied.

Besides, we calculate the PSNR values for the watermarked data. The average PSNR value for the watermarked data is over 80dB with a 32$\times$32-bit watermark, which indicates that the watermarked data is very close to the original data, with minimal distortion introduced by the watermarking process.

These results show the effectiveness of Vol-Mark in achieving low-distortion embedding while preserving the integrity of the original data under attack-free conditions.

\subsection{Conventional and Geometric Attacks}
We evaluate the robustness of the proposed method against both conventional and geometric attacks. 

\textbf{Conventional attacks}: Conventional attacks, such as Gaussian noise, JPEG compression, and median filtering, are commonly used to evaluate a method's robustness against various types of signal distortions. Gaussian noise simulates random perturbations in the data, JPEG compression reduces data quality through data loss, and median filtering smooths the data to reduce noise\cite{tao2014robust}. These tests assess the method's resistance to signal degradation, quality reduction, and noise filtering.

\begin{figure*}[htbp]
    \centering
    \subfloat[No attack]{
        \includegraphics[width=0.48\linewidth]{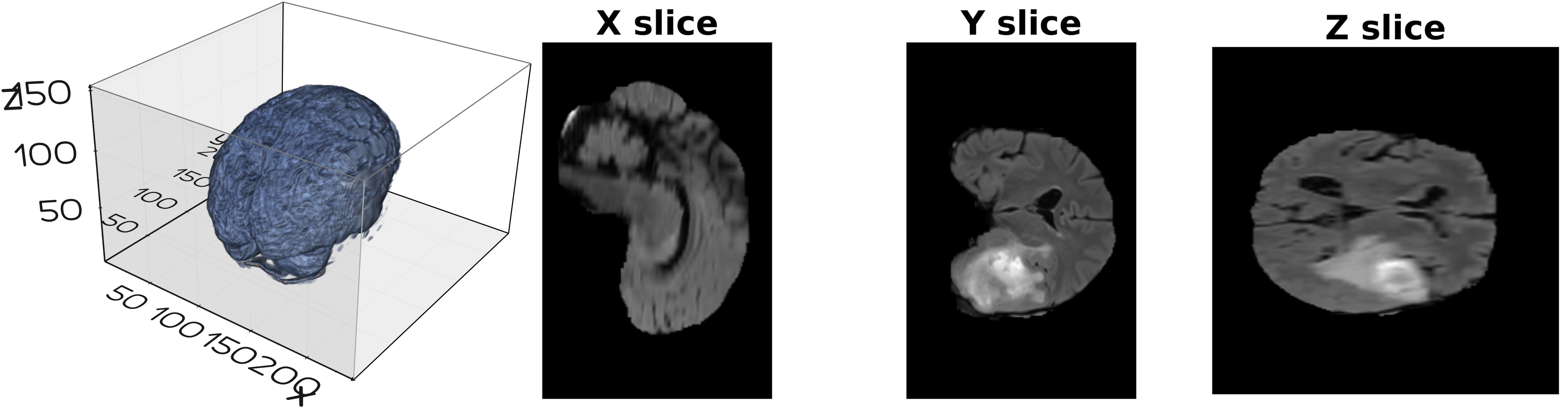}
        \label{fig:noattack}
    }
    \hfill
    \subfloat[Gaussian noise 15\%]{
        \includegraphics[width=0.48\linewidth]{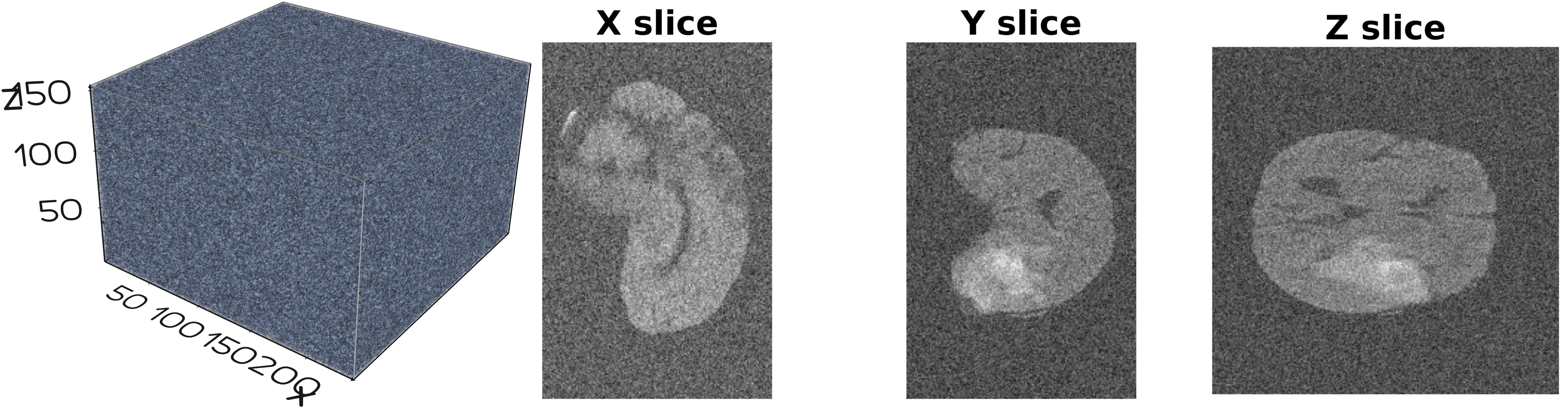}
        \label{fig:gn}
    }
    \par\medskip
    \subfloat[Salt-and-pepper 10\%]{
        \includegraphics[width=0.48\linewidth]{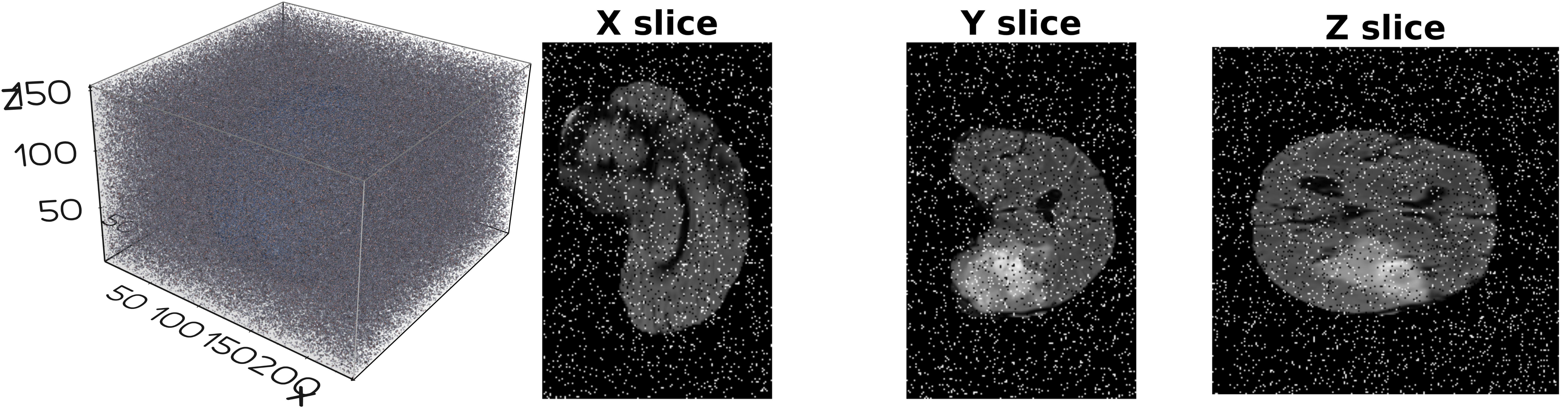}
        \label{fig:spn}
    }
    \hfill
    \subfloat[JPEG Compression 50\%]{
        \includegraphics[width=0.48\linewidth]{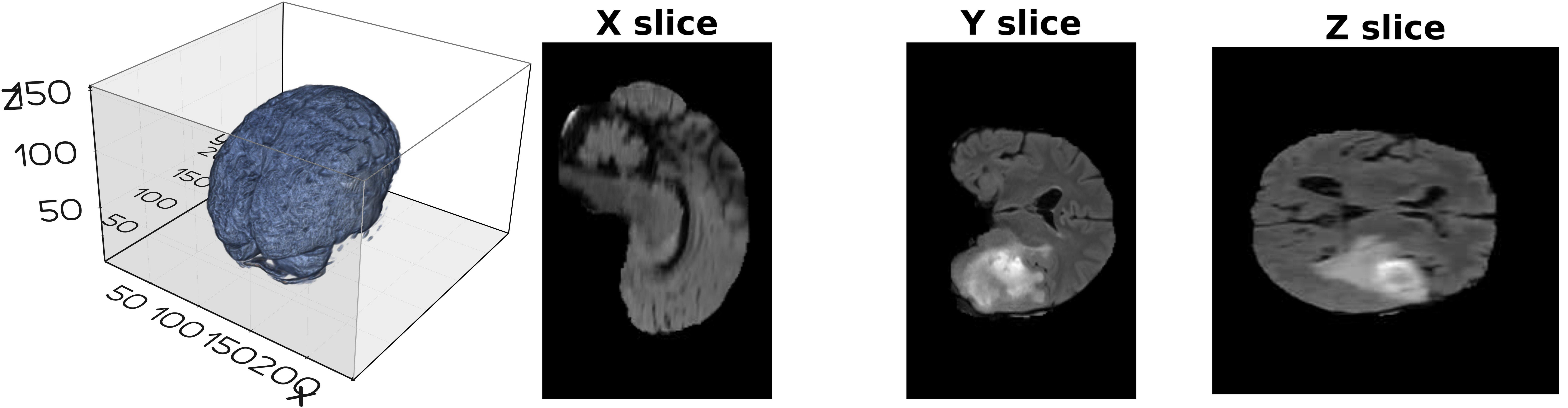}
        \label{fig:jpeg}
    }
    \par\medskip
    \subfloat[Median filtering $5\times5\times5$]{
        \includegraphics[width=0.48\linewidth]{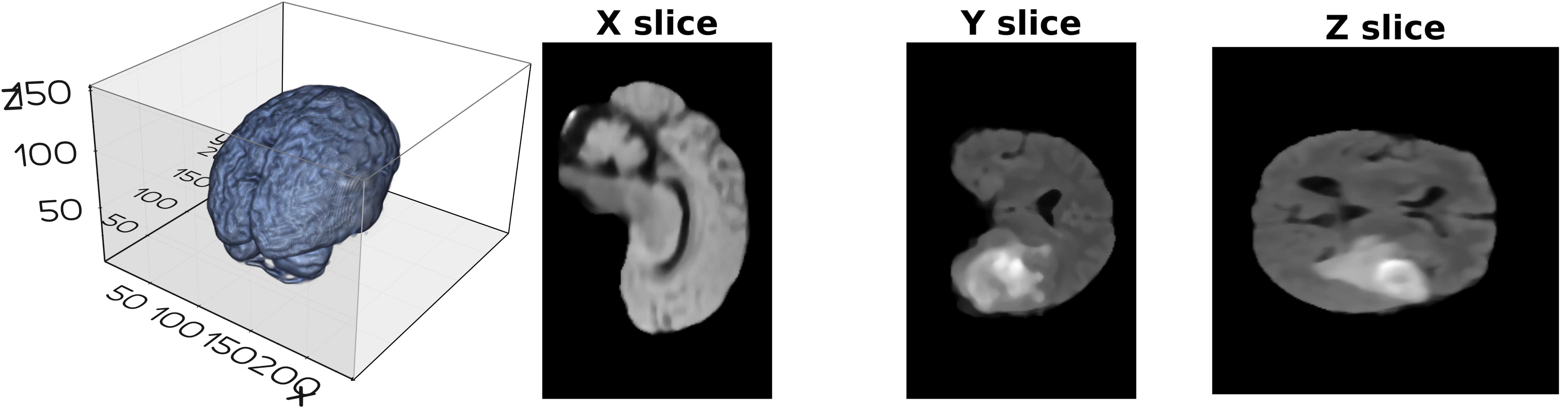}
        \label{fig:mf}
    }
    \hfill
    \subfloat[Average filtering $5\times5\times5$]{
        \includegraphics[width=0.48\linewidth]{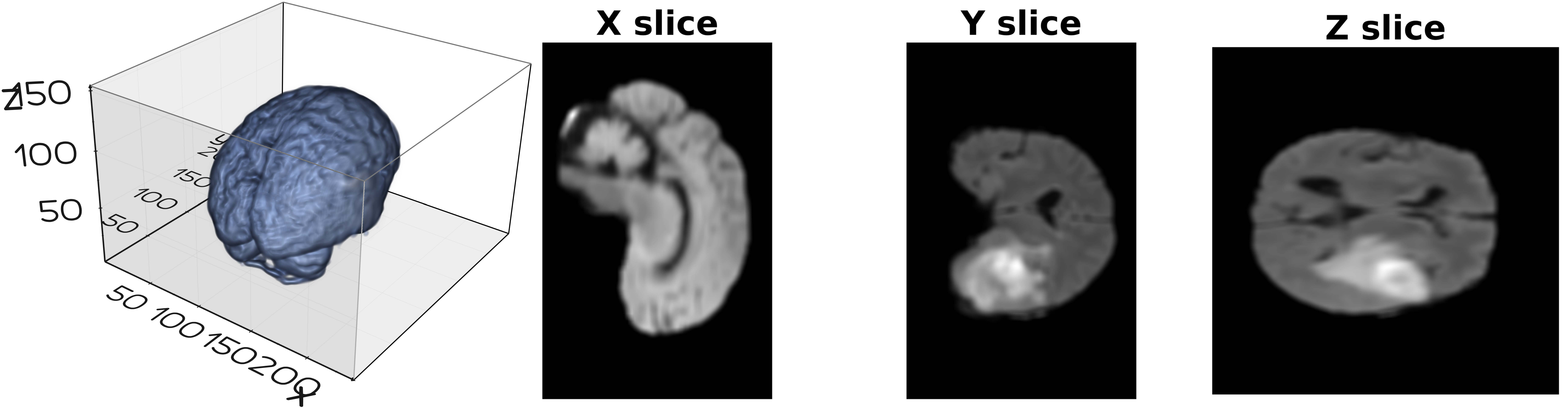}
        \label{fig:gf}
    }
    \par\medskip
    \caption{Visualization of volume data from Task01 (Brain Tumours) after conventional attacks.}
    \label{fig:conven}
\end{figure*}

In our experiments, we applied Gaussian noise with standard deviations of 1\%, 5\%, 10\%, 20\%, and 25\%; Salt-and-pepper noise with the intensities of 1\%, 3\%, 5\%, 10\%, and 15\%; JPEG compression with quality factors of 50, 60, 70, 80, and 90; and median and average filtering with window sizes of 3, 5, and 7 to test the robustness of the medical volume data. 
Fig.~\ref{fig:conven} shows the volume data after conventional attacks.

\begin{figure*}[htbp]
    \centering
    \subfloat[Scaling attack 0.75]{
        \includegraphics[width=0.48\linewidth]{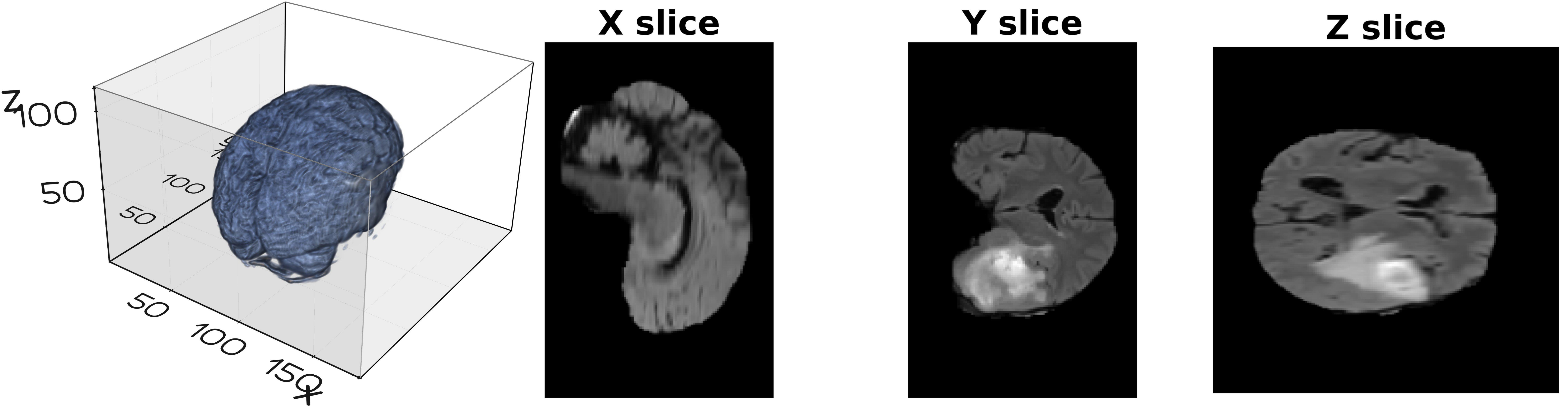}
        \label{fig:sl}
    }
    \hfill
    \subfloat[Cropping attack 10\%]{
        \includegraphics[width=0.48\linewidth]{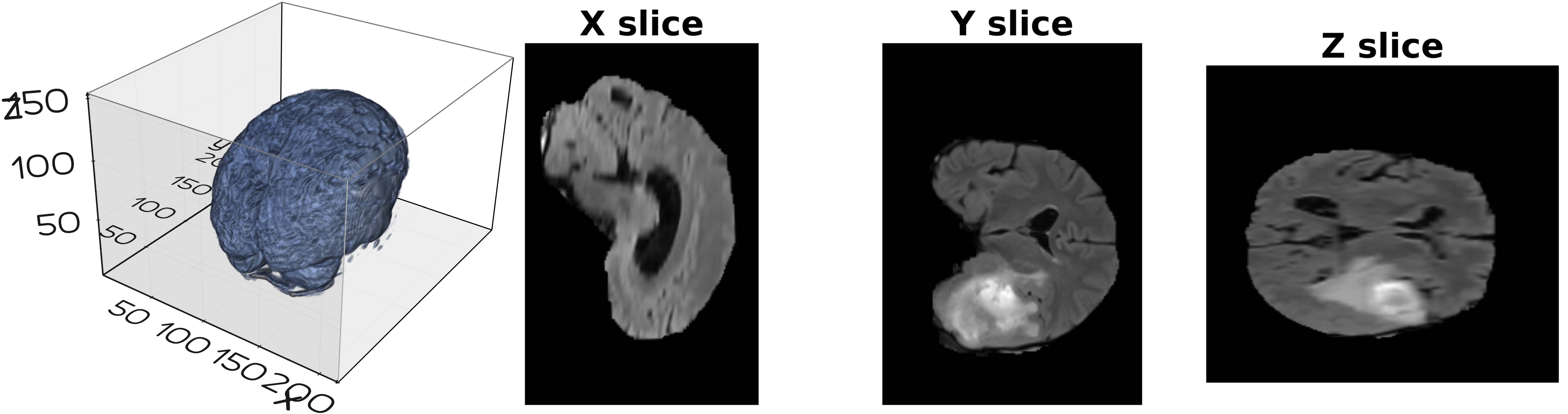}
        \label{fig:cp}
    }
    \par\medskip
    \subfloat[Rotation attack $15^\circ$]{
        \includegraphics[width=0.48\linewidth]{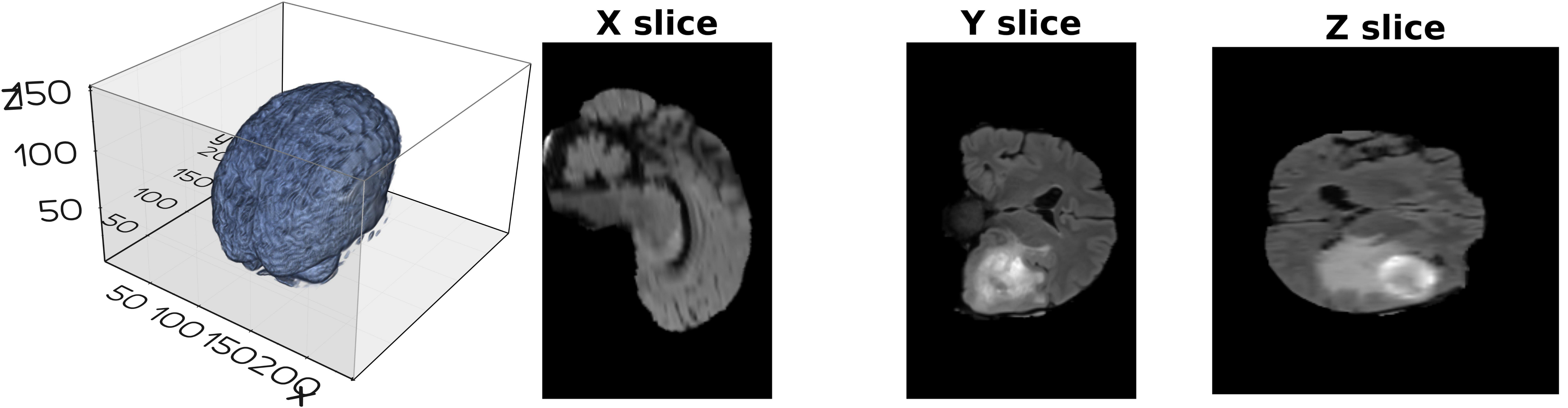}
        \label{fig:rt}
    }
    \hfill
    \subfloat[Translation attack 20\%]{
        \includegraphics[width=0.48\linewidth]{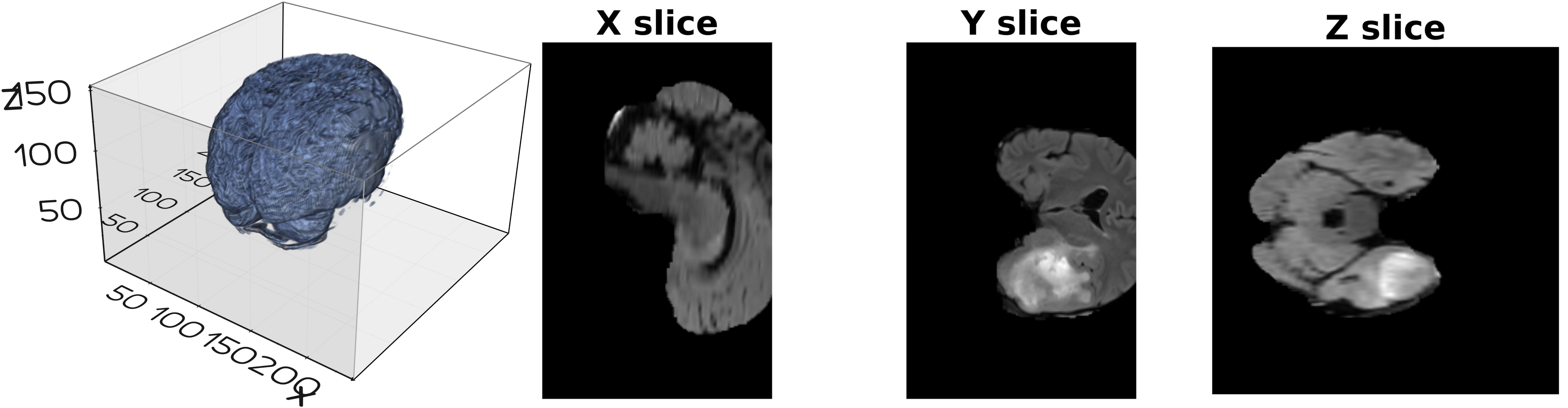}
        \label{fig:tl}
    }
    \par\medskip
    \caption{Visualization of volume data from Task01 (Brain Tumours) after geometric attacks.}
    \label{fig:gmo}
\end{figure*}

\begin{figure*}
    \centering
    \subfloat[Random cropping on Task01 (Brain Tumours)]{
        \includegraphics[width=0.48\linewidth]{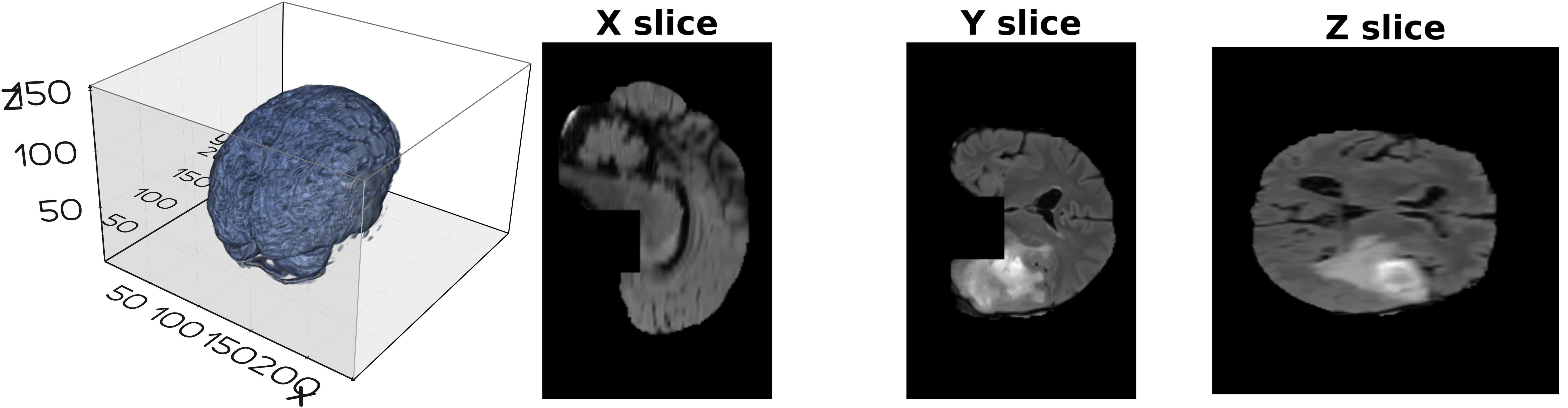}
        \label{fig:rt}
    }
    \hfill
    \subfloat[Random cropping on Task07 (Pancreas)]{
        \includegraphics[width=0.48\linewidth]{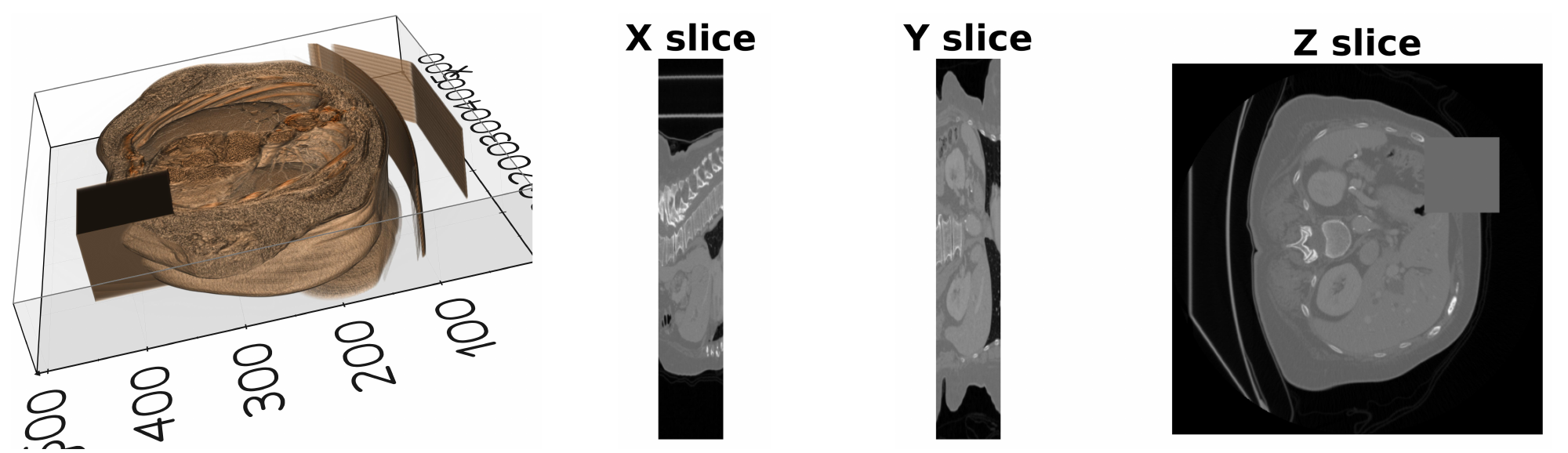}
        \label{fig:tl}
    }
    \par\medskip
    \caption{Visualization of volume data after random cropping attacks (5\%).}
    \label{fig:volrandomcrop}
\end{figure*}

The results in Table \ref{tab:ber1} show that Vol-Mark exhibits strong robustness against conventional attacks, maintaining a high NC close to 0.9 and a low BER below 0.10
on both Task01 and Task03 even under strong noise and filtering. Under JPEG compression attack, the method remains highly stable, achieving NC close to 0.99 for reliable watermark extraction.

\textbf{Geometric attacks}: Geometric attacks involve spatial transformations such as rotation, scaling, translation, and cropping. These attacks are designed to test the method's resilience to changes in angular orientation, size variation, positional shifts, and partial data loss, which are common in real-world scenarios where volume data might be altered intentionally or unintentionally\cite{gmo}. Such manipulations challenge the method's ability to accurately recover information from distorted or altered data.

In our experiments, we applied scaling attacks with scaling factors of 0.5, 0.75, 1.25, and 1.50, cropping attacks with 2\%, 5\%, and 10\%, rotation attacks with rotation angles of $1^\circ$, $5^\circ$, $10^\circ$, and $15^\circ$ in YZ plane, and translation attacks with 1\%, 5\%, and 10\% along the Z-axis to the medical volume data. 
Fig.~\ref{fig:gmo} shows the volume data after geometric attacks.

The experimental results for these attacks are presented in 
Table~\ref{tab:ber2}.
Vol-Mark still maintains high NC above 0.90 and low BER under scaling, rotation, and translation attacks, with nearly perfect performance under slight translation, 
indicating that Vol-Mark
achieves strong robustness against geometric attacks.

\begin{table*}[!htbp]
\caption{NC and BER results under geometric attacks. PSNR not reported for scaling/cropping due to size changes.}
\begin{center}
\begin{tabular}{c|c|ccc|ccc|ccc}
\toprule
\multirow{2}{*}{\textbf{Types of attacks}} & \multirow{2}{*}{\textbf{Intensity}} 
  & \multicolumn{3}{c}{\textbf{Task01} (Brain Tumours)}& \multicolumn{3}{c}{\textbf{Task07} (Pancreas)}&  \multicolumn{3}{c}{Average}\\ 
\cline{3-11}& & PSNR$\uparrow$& BER$\downarrow$ &NC$\uparrow$& PSNR$\uparrow$& BER$\downarrow$  &NC$\uparrow$& PSNR$\uparrow$& BER$\downarrow$  &NC$\uparrow$\\ 
\midrule

\multirow{4}{*}{Scaling} 
& 0.5& /& 0.0155 &0.9803 & /& 0.0195 &0.9754 & /& 0.0175 &0.9779 
\\ 
 & 0.75 
& /& 0.0084 &0.9894 & /& 0.0110 &0.9861 & /& 0.0097 &0.9878 
\\ 
 & 1.25& /& 0.0048 &0.9938 & /& 0.0065 &0.9917 & /& 0.0057 &0.9928 
\\ 
 & 1.5& /& 0.0043 &0.9945 & 
/& 
0.0061 &0.9923 & /& 0.0052 &0.9934 
\\ 
\midrule
\multirow{3}{*}{Cropping} 
& 2\%& /& 0.0126 &0.9840 & /& 0.0206 &0.9739 & /& 0.0166 &0.9790 
\\ 
 & 5\%& /& 0.0265 &0.9666 & /& 0.0531 &0.9336 & /& 0.0398 &0.9501 
\\ 
 & 10\%& /& 0.0502 &0.9372 & /& 0.0855 &0.8940 & /& 0.0679 &0.9156 
\\ 
  
\midrule
\multirow{4}{*}{Rotation} 
& $1^\circ$& 37.77 & 0.0057 &0.9928 & 
25.14 & 
0.0211 &0.9733 & 31.46 & 0.0134 &0.9831 
\\ 
& $5^\circ$& 27.73 & 0.0258 &0.9674 & 19.67 & 0.0476 &0.9403 & 23.70 & 0.0367 &0.9539 
\\ 
& $10^\circ$& 24.55 & 0.0486 &0.9393 & 
17.38 & 
0.0745 &0.9074 & 20.97 & 0.0616 &0.9234 
\\ 
& $15^\circ$& 22.93 & 0.0712 &0.9116 & 16.12 & 0.1003 &0.8760 & 19.53 & 0.0858 &0.8938 
\\ 
\midrule

\multirow{3}{*}{Translation} 
& 1\%& 47.29 & 0.0000 &1.0000 & 44.74 & 0.0000 &1.0000 & 46.02 & 0.0000 &1.0000 
\\ 
& 5\%& 23.87 & 0.0359 &0.9548 & 22.12 & 0.0597 &0.9254 & 23.00 & 0.0478 &0.9401 
\\
& 10\%& 20.98 & 0.0687 &0.9144 & 19.35 & 0.0913 &0.8872 & 20.17 & 0.0800 &0.9008 
\\
\bottomrule
\end{tabular}
\label{tab:ber2}
\end{center}
\end{table*}

\textbf{Hybrid attacks}: 
We further evaluate the performance of Vol-Mark under hybrid attacks, where multiple attacks are applied simultaneously. In our experiments, we construct several hybrid attack scenarios by combining two typical attacks, including JPEG compression with Gaussian noise, median filtering with salt-and-pepper noise, rotation with scaling, translation with rotation, and cropping with Gaussian noise. Different intensity levels are used for each operation to simulate practical distortions encountered in real-world transmission and processing. As summarized in Table~\ref{tab:hybrid_ber}, Vol-Mark preserves strong robustness and can reliably extract the embedded watermark even under hybrid attacks.
It consistently achieves high NC values across all attack combinations, with an average NC above 0.91. Task01 performs better than Task07, likely because pancreas CT scans occupy a larger proportion of the background with more complex structures, which are more sensitive to distortions such as rotation, making stable watermark extraction more challenging. Nevertheless, Vol-Mark still presents strong performance on Task07, maintaining NC values around 0.88–0.99 and BER values between 0.008 and 0.10, indicating reliable robustness even in more complex scenarios

\begin{table*}[!htbp]
\caption{NC and BER results under hybrid attacks.}
\begin{center}
\begin{tabular}{c|c|ccc|ccc|ccc}
\toprule
\multirow{2}{*}{\textbf{Types of attacks}} & \multirow{2}{*}{\textbf{Intensity}} 
  & \multicolumn{3}{c}{\textbf{Task01} (Brain Tumours)}& \multicolumn{3}{c}{\textbf{Task07} (Pancreas)}&  \multicolumn{3}{c}{Average}\\ 
\cline{3-11}& & PSNR$\uparrow$& BER$\downarrow$ &NC$\uparrow$& PSNR$\uparrow$& BER$\downarrow$  &NC$\uparrow$& PSNR$\uparrow$& BER$\downarrow$  &NC$\uparrow$\\ 
\midrule

\multirow{2}{*}{\shortstack{JPEG compression and \\ Gaussian noise}} 
& 70, 1\%& 35.42 & 0.0054 &0.9931 & 33.64 & 0.0082 &0.9896 & 34.53 & 0.0068 &0.9914 
\\ 
 & 50, 5\% 
& 23.57 & 0.0136 &0.9828 & 23.75 & 0.0175 &0.9778 & 23.66 & 0.0156 &0.9803 
\\ 
\midrule

\multirow{2}{*}{\shortstack{Median filtering and \\ Salt-and-pepper Noise}} 
& 3, 1\%& 24.54 & 0.0231 &0.9708 & 25.10 & 0.0345 &0.9566 & 24.82 & 0.0288 &0.9637 
\\ 
 & 5, 5\% 
& 18.63 & 0.0551 &0.9313 & 20.51 & 0.0582 &0.9274 & 19.57 & 0.0567 &0.9294 
\\ 
\midrule
  
\multirow{2}{*}{Rotation and Scaling} 
& $5^\circ$, 0.75& /& 0.0261 &0.9671 & /& 0.0472 &0.9408 & /& 0.0367 &0.9540 
\\ 
 & $10^\circ$, 1.25 
& /& 0.0485 &0.9394 & /& 0.0747 &0.9071 & /& 0.0616 &0.9233 
\\ 
\midrule

\multirow{2}{*}{Translation and Rotation} 
& 3\%, $5^\circ$& 27.59 & 0.0275 &0.9654 & 19.55 & 0.0508 &0.9365 & 23.57 & 0.0392 &0.9510 
\\ 
 & 5\%, $10^\circ$ 
& 23.99 & 0.0562 &0.9300 & 16.95 & 0.0973 &0.8798 & 20.47 & 0.0768 &0.9049 
\\ 
\midrule

\multirow{2}{*}{Cropping and Gaussian Noise} 
& 5\%, 1\%& /& 0.0272 &0.9657 & /& 0.0533 &0.9333 & /& 0.0403 &0.9495 
\\ 
 & 10\%, 5\% 
& /& 0.0508 &0.9364 & /& 0.0859 &0.8936 & /& 0.0684 &0.9150 
\\ 

\bottomrule
\end{tabular}
\label{tab:hybrid_ber}
\end{center}
\end{table*}

In conclusion, the experimental results show that our proposed method exhibits strong robustness against various conventional, geometric and their hybrid attacks.

\subsection{Comparison with Existing Methods}
\label{sec:compairison}
We evaluate the performance of our proposed method in comparison with existing approaches 3D-DTCWT~\cite{3dzero1} and ADCL-ZW~\cite{liu2025attack}. All experiments were conducted under the same conditions and on the same dataset to ensure fairness. 
In the implementation, the batch size is set to 16 and the training dataset contains 150 samples. Therefore, training for 200 epochs results in approximately 2000 iterations. For ADCL-ZW, the experimental settings generally follow those reported in~\cite{liu2025attack}. To maintain comparable training conditions, we use the same batch size and set the number of training iterations to 2000.

\begin{figure*}[!htbp]
    \centering
    \includegraphics[width=\linewidth]{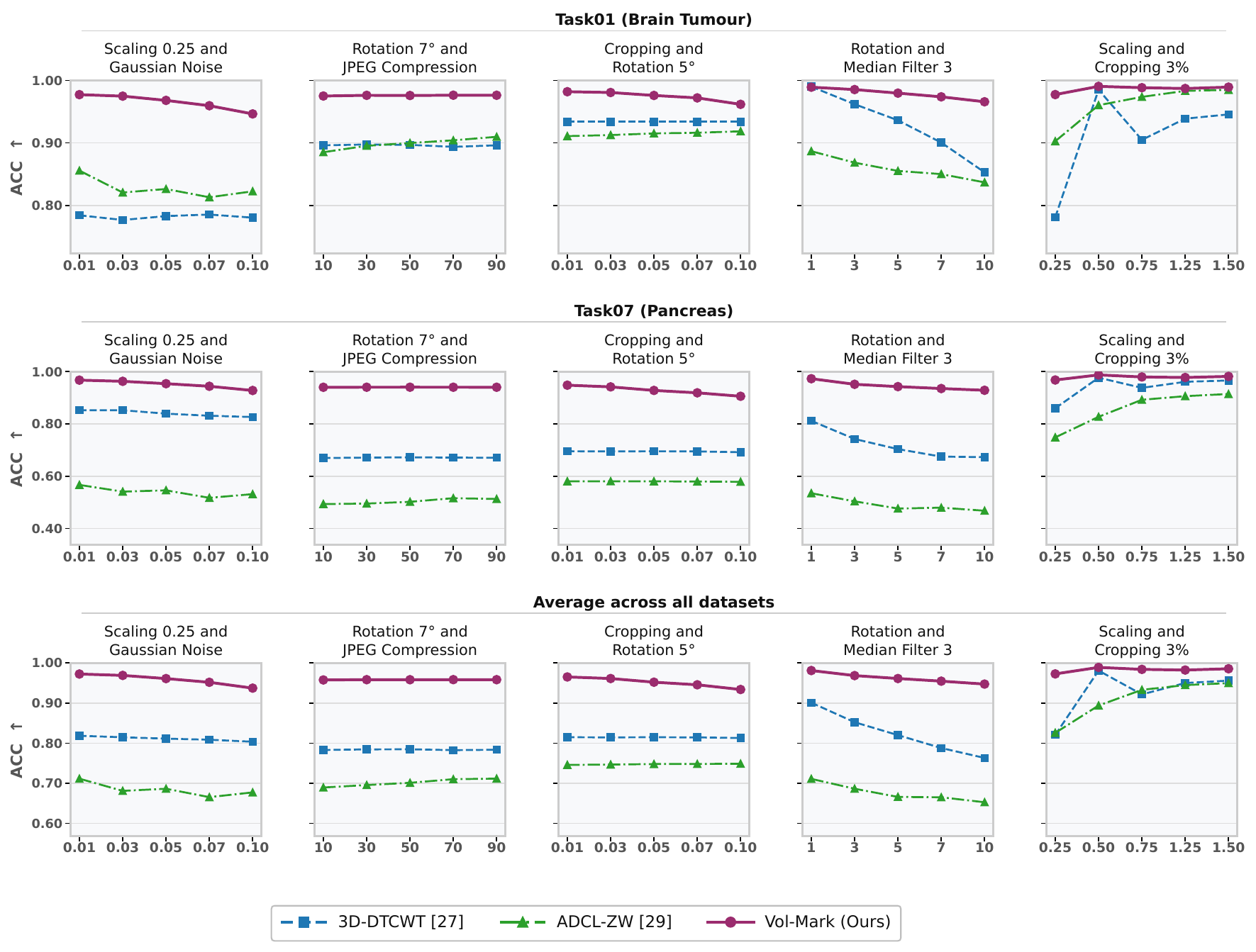}
    \caption{Accuracy comparison of Vol-Mark and baselines  under hybrid attacks.}
    \label{fig:compair1}
\end{figure*}

We conduct comparative experiments under five attacks: Gaussian noise, JPEG compression, Z-axis cropping, rotation in YZ plane, and scaling. To enhance the attack intensity, each attack is supplemented with an additional attack: Gaussian noise with scaling at 0.25; JPEG compression with $7^\circ$ rotation; cropping with a $5^\circ$ rotation; rotation with median filter of kernel size 3; and scaling with cropping ratio of 0.03.
The ACC results (i.e., 1-BER) are presented in Fig.~\ref{fig:compair1}. Vol-Mark is the only method that consistently achieves ACC above 0.90 across all attack scenarios on both Task01 and Task07. In contrast, the compared methods exhibit varying degrees of performance degradation as attack intensity increases.
When subjected to aggressive scaling and rotation attacks, the compared methods suffer significant ACC degradation, while Vol-Mark remains stable, highlighting its superiority in handling geometric attacks. Furthermore, the performance advantage of Vol-Mark is more evident on Task07 (Pancreas), where the compared methods consistently yield lower ACC values, with ADCL-ZW falling below 0.50 under several attack conditions.
These results show that Vol-Mark achieves superior robustness compared to the other baselines under hybrid attack scenarios.

\begin{table}[!htbp]
\caption{Results under conventional attacks on Task03.}
\begin{center}
\begin{tabular}{c|c|ccc}
\toprule
\multirow{2}{*}{\textbf{Types of attacks}} & \multirow{2}{*}{\textbf{Intensity}} 
  & \multicolumn{3}{c}{\textbf{Task03} (Liver)}\\ 
\cline{3-5}
 & & PSNR$\uparrow$& BER$\downarrow$ &NC$\uparrow$\\ 
\midrule

\multirow{5}{*}{Gaussian noise} 
& 1\% & 45.00 & 0.0000 &1.0000 
\\ 
 & 5\% & 36.93 & 0.0018 &0.9977 
\\ 
 & 10\% & 23.70 & 0.0121 &0.9846 
\\ 
 & 20\% & 17.71 & 
0.0263 &0.9667 
\\ 
 & 25\% & 
14.19 & 0.0390 &0.9509 
\\ 
\midrule
\multirow{5}{*}{Salt-and-pepper noise} 
& 1\% & 9.76 & 0.0635 &0.9209 
\\ 
 & 3\% & 22.75 & 0.0125 &0.9842 
\\ 
 & 5\% & 17.97 & 0.0233 &0.9705 
\\ 

 & 10\% & 15.78 & 0.0353 &0.9555 
\\ 

 & 15\% & 12.79 & 0.0644 &0.9205 
\\ 

\midrule
\multirow{5}{*}{JPEG Compression} 
& 50\%& 
11.01 & 
0.0901 &0.8901 
\\ 
& 60\%& 34.61 & 0.0046 &0.9941 
\\
& 70\%& 
35.36 & 
0.0042 &0.9947 
\\ 
& 80\%& 
36.39 & 0.0038 &0.9952 
\\ 
& 90\%& 37.92 & 
0.0040 &0.9950 
\\ 
\midrule
\multirow{2}{*}{Median filtering} 
& 3 & 
40.53 & 0.0040 &0.9949 
\\ 
& 5& 30.64 & 0.0286 &0.9640 
\\
& 7& 28.13 & 0.0441 &0.9447 
\\
\midrule
\multirow{2}{*}{Average filtering} 
& 3 & 
27.11 & 0.0593 &0.9261 
\\
& 5& 29.68 & 0.0089 &0.9887 
\\
& 7& 27.40 & 0.0213 &0.9731 
\\

\bottomrule
\end{tabular}
\label{tab:abber1}
\end{center}
\end{table}

\begin{table}[!htbp]
\caption{Results under geometric attacks on Task03.}
\begin{center}
\begin{tabular}{c|c|ccc}
\toprule
\multirow{2}{*}{\textbf{Types of attacks}} & \multirow{2}{*}{\textbf{Intensity}} 
  & \multicolumn{3}{c}{\textbf{Task03} (Liver)}\\ 
\cline{3-5}& & PSNR$\uparrow$& BER$\downarrow$ &NC$\uparrow$\\ 
\midrule

\multirow{4}{*}{Scaling} 
& 0.5& /& 0.0128 &0.9837 
\\ 
 & 0.75 
& /& 0.0071 &0.9910 
\\ 
 & 1.25& /& 0.0039 &0.9950 
\\ 
 & 1.5& /& 
0.0039 &0.9950 
\\ 
\midrule
\multirow{3}{*}{Cropping} 
& 2\%& /& 0.0223 &0.9718 
\\ 
 & 5\%& /& 0.0584 &0.9273 
\\ 
 & 10\%& /& 0.0876 &0.8918 
\\ 
  
\midrule
\multirow{4}{*}{Rotation} 
& $1^\circ$& 
26.19 & 
0.0173 &0.9781 
\\ 
& $5^\circ$& 20.54 & 0.0476 &0.9405 
\\ 
& $10^\circ$& 
18.28 & 
0.0880 &0.8917 
\\ 
& $15^\circ$& 
17.04 & 0.1123 &0.8624 
\\ 
\midrule

\multirow{3}{*}{Translation} 
& 1\%& 
45.00 & 0.0000 &1.0000 
\\ 
& 5\%& 21.86 & 0.0516 &0.9354 
\\
& 10\%& 19.46 & 0.0909 &0.8880 
\\

\bottomrule
\end{tabular}
\label{tab:abber2}
\end{center}
\end{table}

\begin{table}[!htbp]
\caption{Results under hybrid attacks on Task03.}
\begin{center}
\begin{tabular}{c|c|ccc}
\toprule
\multirow{2}{*}{\textbf{Types of attacks}} & \multirow{2}{*}{\textbf{Intensity}} 
  & \multicolumn{3}{c}{\textbf{Task03} (Liver)}\\ 
\cline{3-5}& & PSNR$\uparrow$& BER$\downarrow$ &NC$\uparrow$\\ 
\midrule

\multirow{2}{*}{\shortstack{JPEG compression and \\ Gaussian noise}} 
& 70, 1\%& 33.96 & 0.0046 &0.9941 
\\ 
 & 50, 5\% 
& 23.56 & 0.0143 &0.9819 
\\ 
\midrule

\multirow{2}{*}{\shortstack{Median filtering and \\ Salt-and-pepper Noise}} 
& 3, 1\%& 24.57 & 0.0320 &0.9597 
\\ 
 & 5, 5\% 
& 20.66 & 0.0531 &0.9336 
\\ 
\midrule
  
\multirow{2}{*}{Rotation and Scaling} 
& $5^\circ$, 0.75& /& 0.0488 &0.9392 
\\ 
 & $10^\circ$, 1.25 
& /& 0.0882 &0.8915 
\\ 
\midrule

\multirow{2}{*}{Translation and Rotation} 
& 3\%, $5^\circ$& 20.32 & 0.0462 &0.9420 
\\ 
 & 5\%, $10^\circ$ 
& 17.75 & 0.0864 &0.8929 
\\ 
\midrule

\multirow{2}{*}{Cropping and Gaussian Noise} 
& 5\%, 1\%& /& 0.0584 &0.9273 
\\ 
 & 10\%, 5\% 
& /& 0.0876 &0.8918 
\\ 

\bottomrule
\end{tabular}
\label{tab:abhybrid_ber}
\end{center}
\end{table}

\section{Ablation Study}
\label{sec:ablation}
\subsection{Vol-Mark on Larger Dataset}
As shown in Table~\ref{tab:msd}, Task03 contains significantly larger data volumes than Task01 and Task07. Moreover, the data sizes in Task03 vary considerably across samples, with the minimum shape being [512, 79, 42] and the maximum shape reaching [512, 512, 1026]. In contrast, the data sizes in the other two datasets are relatively balanced. To further evaluate the scalability of Vol-Mark, we report its performance on Task03 under various attacks. Table~\ref{tab:abber1} presents the results under conventional attacks, while Table~\ref{tab:abber2} and Table~\ref{tab:abhybrid_ber} report the results under geometric and hybrid attacks, respectively.
In Task03, Vol-Mark has high NC values typically above 0.94 under most conventional attacks and up to 0.99 under mild distortions, with BER generally below 0.05. Compared with Task01 and Task07, Task03 achieves comparable or slightly better performance under several attacks, especially under mild noise and compression. Under geometric attacks, NC remains above 0.89 in most cases, which is consistent with the other tasks. Even under hybrid attacks, NC stays around 0.89–0.99, further indicating that Vol-Mark achieves strong robustness across tasks with different data shapes.

\begin{figure*}[htbp]
    \centering
    \includegraphics[width=\linewidth]{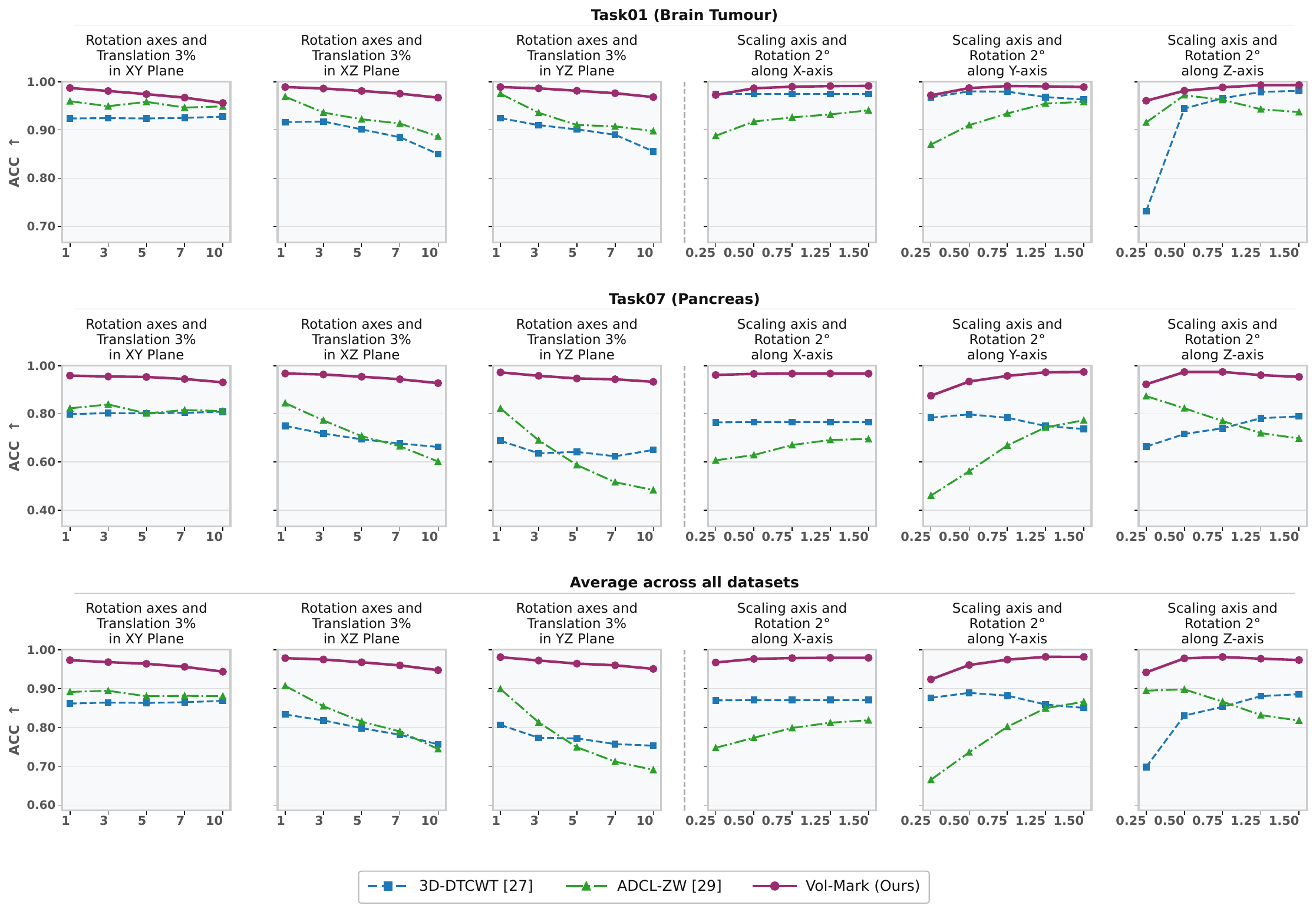}
    \caption{Accuracy comparison of Vol-Mark and baselines under 3D attacks.
    }
    \label{fig:3dscaling}
\end{figure*}

\subsection{Vol-Mark under 3D Attacks}
Vol-Mark shows superior performance compared with the existing methods under 2D geometric attacks as shown in Fig.~\ref{fig:compair1}.
Unlike 2D data, 3D data can undergo transformations along three spatial axes, which introduces additional challenges for robust watermark. To further investigate the robustness of Vol-Mark under different directional transformations, we conduct experiments with two types of geometric attacks, including scaling and rotation, along each spatial axis. Specifically, the attacks are applied independently along the X-, Y-, and Z-axes to evaluate the directional robustness of the methods. Extra attacks are applied to enhance experiment intensity: XY plane rotation of $2^\circ$ on scaling and Z-axis translation of 3\% on rotation attacks. The experimental results are illustrated in Fig.~\ref{fig:3dscaling}. It can be observed that Vol-Mark consistently maintains ACC above 0.90 across all rotation axes and scaling conditions, while 3D-DTCWT and ADCL-ZW exhibit significant ACC degradation under out-of-plane rotations and Y/Z-axis scaling. This is particularly evident on Task07, where the ACC of ADCL-ZW drops to approximately 0.50 under YZ-plane rotation and shows a consistent downward trend under Z-axis scaling regardless of the scaling factor. The inferior performance of the compared methods can be attributed to their limited 3D feature representations. 3D-DTCWT extracts features from the sign sequences of global low-frequency coefficients via 3D DTCWT-DCT transformation, which remain relatively stable under common signal attacks. However, rotations in the XZ and YZ planes disrupt the inter-slice structure and cause significant changes in the low-frequency coefficient distribution, leading to feature instability and ACC degradation.
For ADCL-ZW, the method uses slice-wise mean values as input, which discards volumetric depth information. Out-of-plane rotations alter the slice composition, leading to significant changes in the slice mean values and consequently degrading feature robustness. 

In contrast, Vol-Mark employs a 3D ResNet-18 trained with contrastive loss to extract volumetric features directly from the entire volume, explicitly capturing inter-slice dependencies and maintaining robustness against both in-plane and out-of-plane geometric transformations.

\subsection{Vol-Mark under Random Cropping}
In the above experiments, the geometric attacks are mainly applied along the three spatial axes. However, in practical data transmission scenarios, packet loss may occur randomly. To simulate this situation, we perform a random cropping experiment by randomly removing a cubic region from the volume data. 

\begin{figure}[t]
    \centering
    \includegraphics[width=\linewidth]{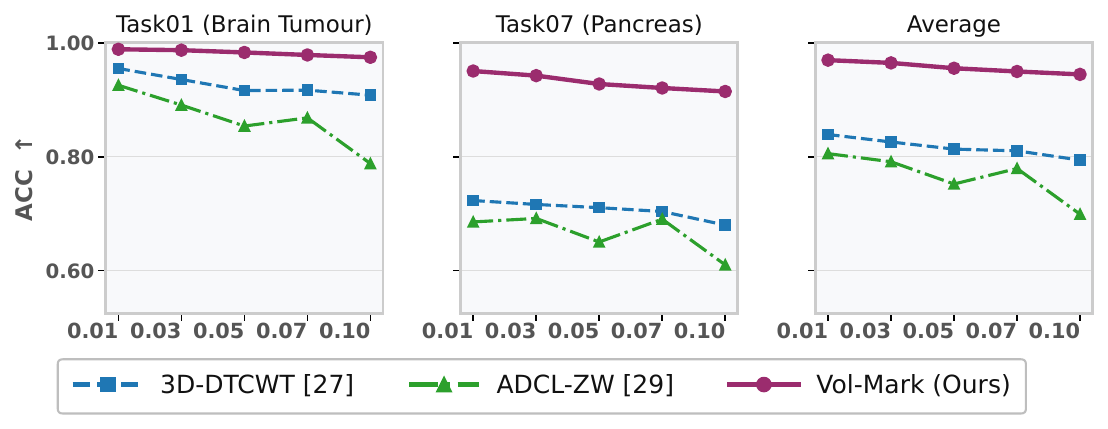}
    \caption{Comparison of accuracy under random cropping.}
    \label{fig:randomcrop}
\end{figure}

To make the simulation more realistic and avoid the influence of background regions, the random packet loss is applied within the ROI. The cropping ratios are set to 1\%, 3\%, 5\%, 7\%, and 10\%. We add an extra rotation attack on Z-axis by $3^\circ$ to further enhance the attack intensity. A sample of the volume data from Task01 and Task07
under a cropping ratio of 5\% are shown in Fig. \ref{fig:volrandomcrop}, and the average experimental results on Task01 and Task07 are presented in Fig. \ref{fig:randomcrop}. Under random cropping attacks, our method Vol-Mark outperforms the compared baselines, achieving an average accuracy improvement of nearly 0.2, which indicates its superior robustness.

\section{Conclusion}
\label{sec:conclusion}
This paper proposed Vol-Mark, a robust reversible-zero watermarking method specifically designed to protect the ownership and authenticity of medical volume data in telemedicine.
Vol-Mark designs a volume feature extractor based on contrastive learning to effectively extract discriminative and stable volumetric features, enhancing robustness under various attacks.
Combined with the encrypted watermark,
Vol-Mark guarantees the security of watermark generation.
Furthermore,
by introducing a novel c-DE technique, Vol-Mark
embeds watermark bits into the expanded differences of neighboring voxels into cubes at low-frequency coefficients, allowing both
low-distortion embedding and lossless data recovery.
Vol-Mark provides both integrity verification and ownership verification, improving the overall reliability of watermark even under data tampering and watermark removal attacks.
Experimental evaluations confirm Vol-Mark achieves high-accuracy reversible embedding under no-attack conditions and shows strong robustness against conventional, geometric, and hybrid attacks.
In particular, Vol-Mark consistently outperforms baselines across almost every hybrid attack scenario, with a more pronounced advantage against 3D attacks.
These results show the potential of Vol-Mark as a reliable solution for ownership protection and integrity assurance in practical medical volume data applications.

\bibliographystyle{IEEEtranS}
\bibliography{ref}

\vfill

\end{document}